\documentclass[prd,twocolumn,superscriptaddress,altaffilletter,showpacs]{revtex4}
\usepackage{amssymb,amsmath}
\usepackage{comment}
\usepackage{graphicx}

\begin{document}

\title{Quasinormal modes of self-dual black holes in loop quantum gravity}
\author{Mehrab Momennia}
\email{mmomennia@ifuap.buap.mx, momennia1988@gmail.com}
\affiliation{Instituto de F\'{\i}sica, Benem\'erita Universidad Aut\'onoma de Puebla,\\
Apartado Postal J-48, 72570, Puebla, Puebla, M\'exico}
\affiliation{Department of Physics, School of Science, Shiraz
University, Shiraz 71454, Iran}
\date{\today }

\begin{abstract}
We study the evolution of a test scalar field on the background geometry of
a regular loop quantum black hole (LQBH) characterized by two loop quantum
gravity (LQG) correction parameters, namely, the polymeric function and the
minimum area gap. The calculations of quasinormal frequencies in
asymptotically flat spacetime are performed with the help of higher-order
WKB expansion and related Pad\'{e} approximants, the improved asymptotic
iteration method (AIM), and time-domain integration. The effects of free
parameters of the theory on the quasinormal modes are studied and deviations
from those of the Schwarzschild BHs are investigated. We show that the LQG
correction parameters have opposite effects on the quasinormal frequencies
and the LQBHs are dynamically stable.
\end{abstract}

\pacs{04.20.Ex, 04.25.Nx, 04.30.Nk, 04.70.-s}
\maketitle

\section{Introduction}

The quasinormal modes (QNMs) are the intrinsic imprints of BH
response to
external perturbations on its background geometry \cite%
{Kokkotas,BertiR,KonoplyaR}. The QNMs spectrum is an essential
characteristic of BHs that depends on BH charges and could be detected
through the gravitational wave interferometers \cite%
{Abbott2016,Abbott2017,Isi}. Hence, this capability allows us to
explore the properties of background spacetime of BHs, check the
validity of the alternative theories of general relativity, and
estimate the BH parameters by studying
gravitational waves (GWs) at the ringdown stage \cite{GWspectroscopy,RoadMap}%
.

Furthermore, some other potent motivations for investigating the
QN oscillations of BHs in different branches of fundamental
physics can be listed as follows. The QNMs spectrum governs the
dynamic stability of BHs
undergoing small perturbations of various test fields \cite%
{Kokkotas,BertiR,KonoplyaR}, the asymptotic behavior of the QN
modes in the flat background plays a crucial role in the
semi-classical approach to quantum gravity \cite{Hod}, the highly
damped QN frequencies used to fix the so-called Barbero-Immirzi
parameter appearing in LQG \cite{Dreyer}, the imaginary part of
the QN frequencies in asymptotically anti-de Sitter spacetime
describes the decay of perturbations of corresponding thermal
state in the conformal field theory \cite%
{Horowitz,LemosAdS,KokkotasAdS,MehrabJHEP,MehrabSultani}, and the
correspondence between the QN frequencies in the eikonal limit and
unstable circular null geodesics that describe the size of the BH
shadow \cite{CardosoUCNG,KonoplyaUCNG,MomenniaPRD}.

On the other hand, scalar fields have been considered extensively
as candidates for dark energy \cite{Gubitosi} and dark matter
\cite{Hu}. They have been also investigated as the inflatons in
the context of cosmology \cite{Cheung}. Background scalar fields
are a generic feature in the string theory \cite{Metsaev,Arvanitaki}, and they have been used
to modify the background spacetime of BHs in the strong-field
regime \cite{Herdeiro,Silva}. Besides, the scalar fields
produce scalar clouds around BHs through superradiant instability \cite%
{Brito}.

In gravitational models non-minimally coupled to scalar fields,
the emitted GWs is a linear combination of GWs in the
gravitational theory and the
scalar field solutions \cite{Tattersall}. Thus, the gravitational waves $%
\bar{h}_{\mu \nu }$ that could potentially be observed, will be a linear
combination of GWs in the gravitational theory, $h_{\mu \nu }$, and the
scalar field solutions of the form%
\begin{equation}
\bar{h}_{\mu \nu }=h_{\mu \nu }+\beta g_{\mu \nu }\Phi ,
\end{equation}%
where $\Phi $\ is the scalar field, $g_{\mu \nu }$\ is the
background metric, and $\beta $ is an arbitrary function of the
scalar field that characterizes the non-minimal coupling. However,
the interaction of spacetime metric and scalar waves depends on
the scalar propagation speed so that interactions are negligible
for luminal scalar waves \cite{Dalang}.

The scalar fields minimally coupled to gravity describe the QNMs
in the context of scalar-tensor theories. More recently, it has
been demonstrated that the Laser Interferometer Space Antenna will
be able to measure the scalar charge with an accuracy of the order
of percent in the extreme mass ratio inspirals \cite{Maselli}.
This analysis indicated that the detectability of the scalar
charge does not depend on the scalar field origin and the
structure of the secondary compact object that is coupled to the
scalar field.

In the extended and modified gravity theories of general
relativity, the QNMs of BH solutions undergoing scalar
perturbations have been
investigated in higher dimensional Einstein-Yang-Mills theory \cite%
{YangMills}, Einstein-Born-Infeld gravity \cite{BornInfeld}, dRGT
massive gravity \cite{dRGT}, conformal Weyl gravity
\cite{MehrabSultani,MomenniaPRD}, and loop quantum gravity
\cite{QNMofLQG}. In addition, the QN modes of Schwarzschild BHs
with Robin boundary conditions \cite{Robin}, the dirty BHs
\cite{Dirty}, the Kaluza--Klein BHs \cite{KKBH}, and charged BHs
with Weyl corrections \cite{Weylcorrections} have been studied.

When it comes to BH physics, the intrinsic singularity inside the
event horizon has a special place. Although the properties of
spacetime outside the event horizon are described by a few
parameters characterizing the BH conserved charges, the curvature
singularity at the center of BHs remained a crucial and
outstanding problem. In this context, people have performed plenty
of efforts to address this issue, such as assigning conformal
symmetry to spacetime, employing nonlinear electrodynamic fields,
and considering quantum corrections to general relativistic
theories. However, we expect a too strong bending of the spacetime
near the BH center such that the general relativity breaks down
and a quantum description of gravity becomes inevitable.

In this paper, we focus on scalar perturbations in the background
spacetime of a non-singular LQBH to investigate the effects of the
LQG correction parameters on the scalar QNM spectrum, explore the
dynamical stability of the BHs, and find deviations from those of
the Schwarzschild solutions. Our regular BH case study, also known
as the self-dual BH, was constructed in the
mini-superspace approach based on the polymerization procedure in LQG \cite%
{LQBH}\ and characterized by the polymeric function and the
minimum area gap as two LQG correction parameters (see
\cite{Perez,Barrau}\ for review papers on BHs in LQG and
\cite{EHDestruction} for the role of quantum corrections on the
destruction of the event horizon). Particle creation by these
LQBHs is investigated and it was shown that the evaporation time
is infinite \cite{ParticleCreation}. The gravitational lensing by
the LQBHs in the strong and weak deflection regimes is studied
\cite{Lensing}. These quantum-corrected BHs were generalized to
axially symmetric spacetimes and their shadow is investigated \cite%
{LQBHJamil}. However, the QNMs of the static case have been
calculated with some defects in
\cite{LQBHJamil,Chen2011,BarrauUniverse}, and we shall address
this issue in the present study as well.

The outline of this paper is as follows. Section \ref{LQBH} is
devoted to a brief review of LQG-corrected BHs and perturbation
equations of a test scalar field. Then, we briefly explain the
higher-order WKB approximation and related Pad\'{e} approximants,
the improved AIM, and the time-domain integration that are used to
investigate the QN modes. In Sec. \ref{QNMs}, we calculate the
QNMs of LQBHs, study the effects of LQG correction parameters on
the QNMs spectrum, and find deviations from those of the
Schwarzschild BHs. Besides, we investigate the dynamical stability
of the LQBHs, and compute the QNMs by employing the higher-order
WKB formula and related Pad\'{e} approximants as a semi-analytic
method. We finish our paper with some concluding remarks.

\section{Loop Quantum Black holes and perturbation equations \label{LQBH}}

The effective LQG-corrected line element, also known as the self-dual
spacetime, with spherical symmetry that is geodesically complete is given by
\cite{LQBH}
\begin{equation}
ds^{2}=-f(r)dt^{2}+\frac{dr^{2}}{g(r)}+h(r)d\Omega ^{2},  \label{metric}
\end{equation}%
where $d\Omega ^{2}$ is the line element of a $2$-sphere and the metric
functions $f(r)$, $g(r)$, and $h(r)$\ can be written as%
\begin{equation}
f(r)=\frac{\left( r-r_{+}\right) \left( r-r_{-}\right) }{r^{4}+A_{0}^{2}}%
\left( r+r_{0}\right) ^{2},  \label{f}
\end{equation}%
\begin{equation}
g(r)=\frac{\left( r-r_{+}\right) \left( r-r_{-}\right) }{r^{4}+A_{0}^{2}}%
\frac{r^{4}}{\left( r+r_{0}\right) ^{2}},  \label{g}
\end{equation}%
\begin{equation}
h(r)=r^{2}+\frac{A_{0}^{2}}{r^{2}},  \label{h}
\end{equation}%
with the\ outer (event) horizon $r_{+}=2M/(1+P)^{2}$, the\ inner (Cauchy)
horizon $r_{-}=2MP^{2}/(1+P)^{2}$, and the polymeric function $P=\left(
\sqrt{1+\epsilon ^{2}}-1\right) /\left( \sqrt{1+\epsilon ^{2}}+1\right) $
arising from the geometric quantum effects of LQG. Besides, $A_{0}$ is
related to the minimum area gap of LQG as $A_{0}=A_{\min }/(8\pi)$ and $%
r_{0}=\sqrt{r_{+}r_{-}}=2MP/(1+P)^{2}$. In the aforementioned relations, $M$%
\ is the total mass of the BHs, and $\epsilon $\ denotes a product of the
Immirzi parameter $\gamma $ and the polymeric parameter $\delta $ satisfying $%
\epsilon =\gamma \delta <<1$.

\begin{figure*}[tbh]
\centering
\includegraphics[width=0.4\textwidth]{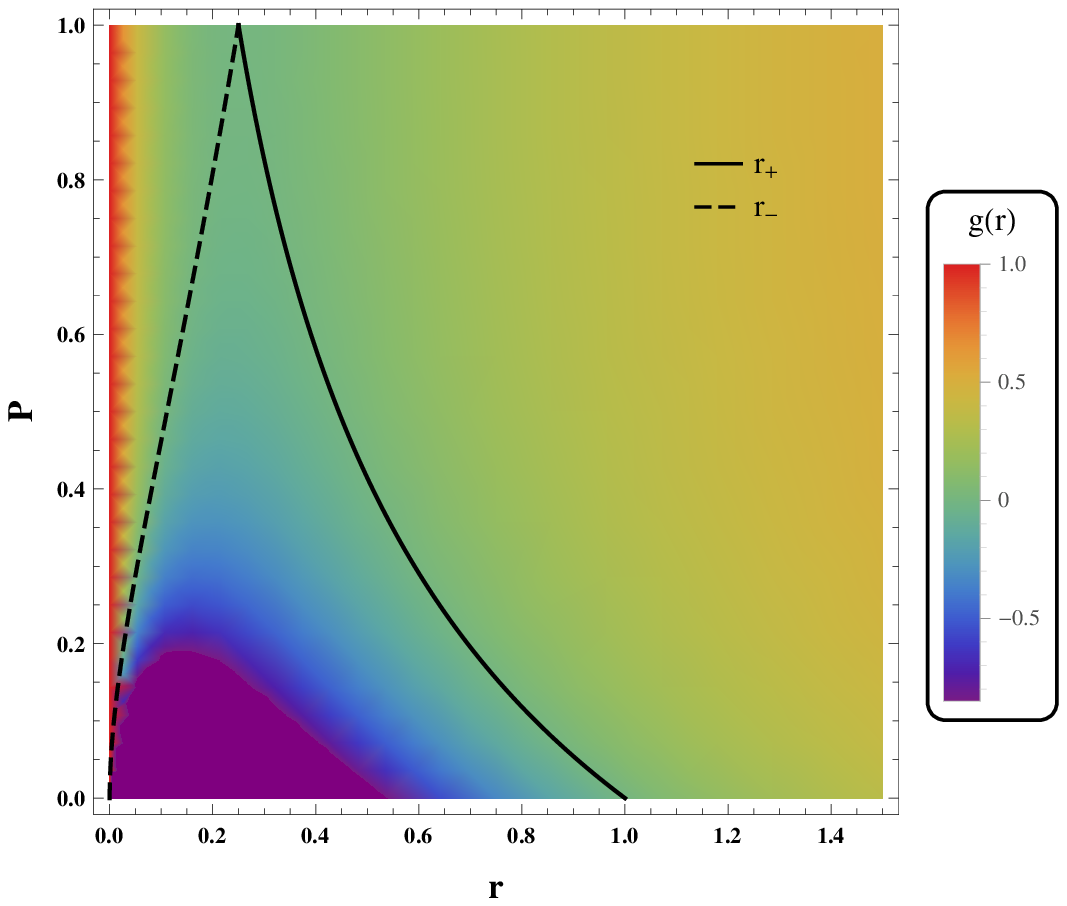} \includegraphics[width=0.4%
\textwidth]{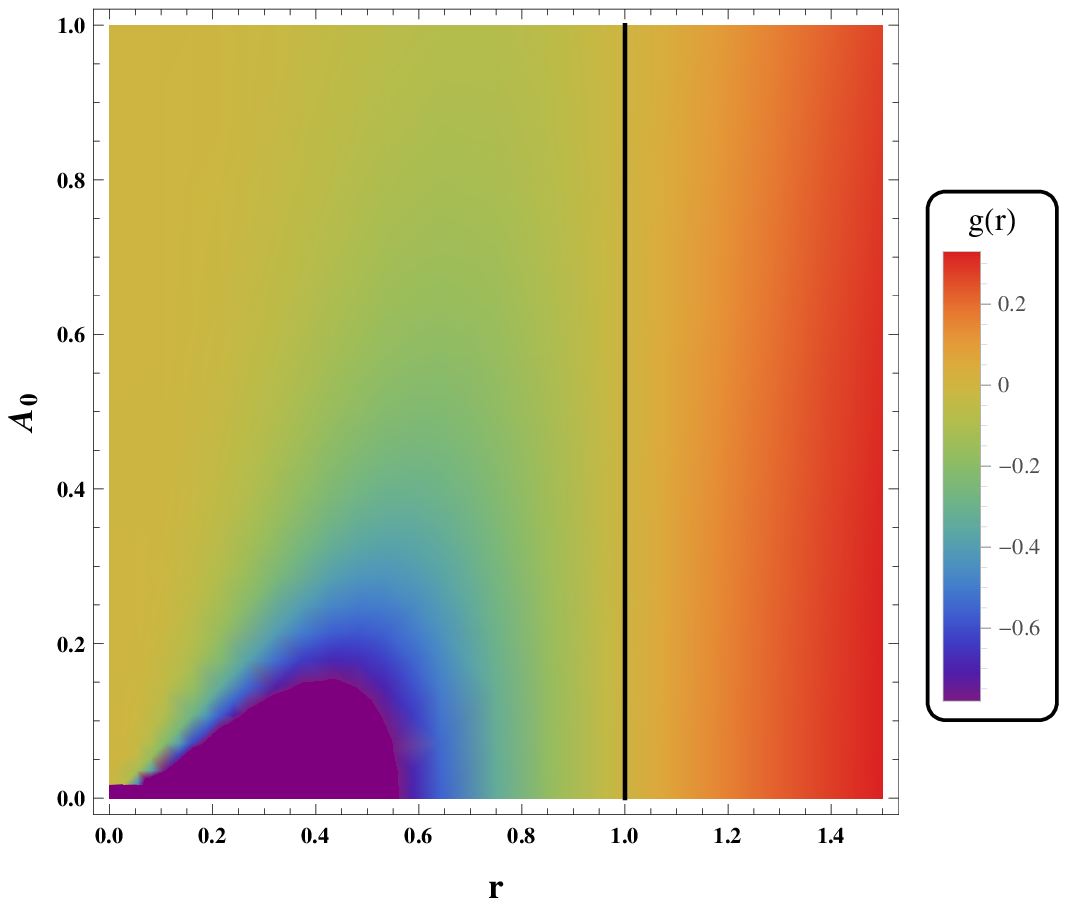}
\caption{The metric function $g(r)$ in $r-P$ plane for $A_{0}=0$ (left
panel) and $r-A_{0}$ plane for $P=0$ (right panel). The non-zero values of
the polymeric function produce an inner horizon and decrease the event
horizon radius (see the left panel). The vertical black line in the right
panel denotes the event horizon radius $r_{+}$ of the single-horizon BH.}
\label{gr}
\end{figure*}

It is worthwhile to mention that the inner horizon is produced due to LQG
generalization, and these BHs reduce to a single-horizon BH whenever the
polymeric function $P$\ vanishes (see Fig. \ref{gr}). Besides, note that the
LQG correction parameters $\epsilon $\ and $A_{0}$\ describe deviations from
the Schwarzschild solutions. Therefore, the LQBHs (\ref{metric}) reduce to
Schwarzschild BHs by taking the limit $\epsilon =0=A_{0}$.
\begin{figure*}[tbh]
\centering
\includegraphics[width=0.4\textwidth]{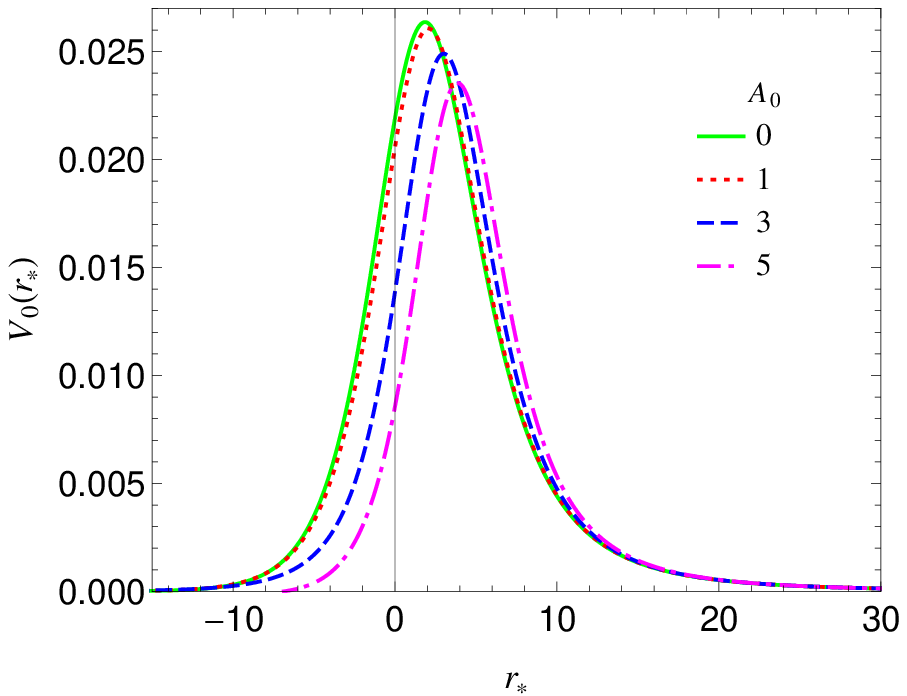} \includegraphics[width=0.4%
\textwidth]{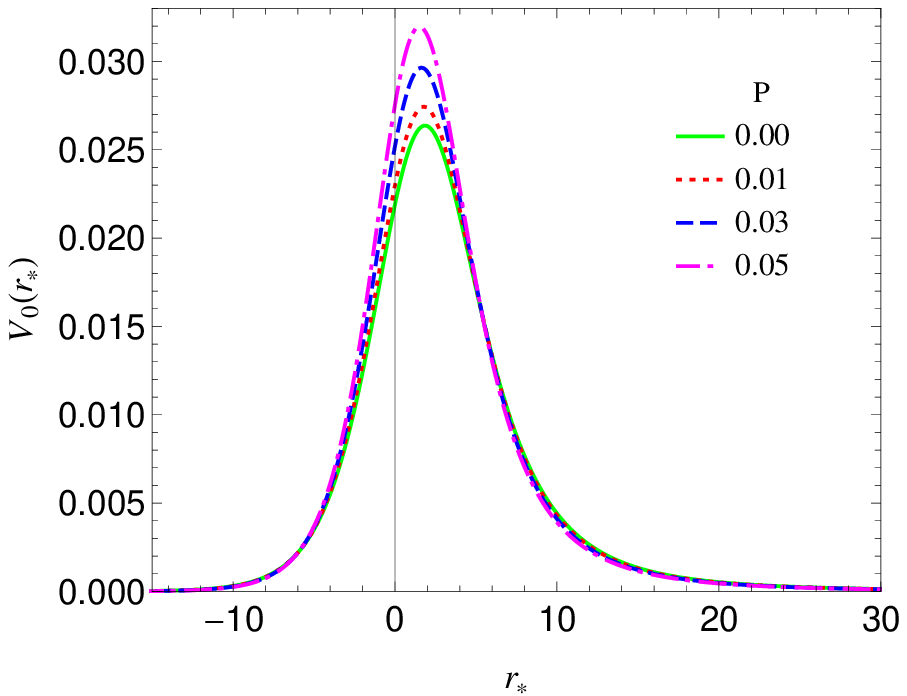}
\caption{The effective potential versus tortoise coordinate for $M=1$, $l=0$%
, $P=0$ (left panel), and $A_{0}=0$ (right panel). The continuous
green curve denotes the Schwarzschild's potential. The potential
forms a barrier and vanishes at both infinities (in order to see
the effects of $A_{0}$ on the potential, large values were
adopted).} \label{Pot}
\end{figure*}

Now, we consider a scalar perturbation in the background of the LQBHs to
investigate their QNMs spectrum. The equation of motion for a minimally
coupled scalar field is given by
\begin{equation}
\nabla _{\mu }\nabla ^{\mu }\Phi \left( t,r,\theta ,\varphi \right) =0.
\label{WEQ}
\end{equation}

The following expansion of modes
\begin{equation}
\Phi \left( t,r,\theta ,\varphi \right) =\int d\omega e^{-i\omega
t}\sum_{l,m}\frac{1}{\sqrt{h(r)}}\Psi _{l}\left( r\right)
Y_{l,m}\left( \theta ,\varphi \right) ,  \label{sh}
\end{equation}%
allows us to find a Schr\"{o}dinger-like wave equation for the radial part $%
\Psi _{l}\left( r\right) $\ of perturbations, and $Y_{l,m}\left( \theta
,\varphi \right) $ denotes the spherical harmonics on a $2$-sphere.
Substituting the decomposition (\ref{sh})\ into the Klein--Gordon equation (%
\ref{WEQ}), the equation of motion reduces to the following wave-like
equation for the radial part of the perturbations
\begin{equation}
\left[ \partial _{r_{\ast }}^{2}+\omega ^{2}-V_{l}\left( r_{\ast }\right) %
\right] \Psi _{l}\left( r_{\ast }\right) =0,  \label{Weq}
\end{equation}
where $\omega$ is the Fourier variable presented in Eq.
(\ref{sh}). In this relation, $V_{l}\left( r_{\ast }\right) $ is
the effective potential that is given by
\begin{eqnarray}
V_{l}\left( r_{\ast }\right) &=&f\left( r\right) \frac{l\left( l+1\right) }{%
h(r)}  \notag \\
&&+\sqrt{\frac{f(r)g(r)}{h(r)}}\partial _{r}\left( \sqrt{f(r)g(r)}\partial
_{r}\sqrt{h(r)}\right) ,  \label{EP}
\end{eqnarray}%
where $l$ is the angular quantum (multipole) number, and$\ r_{\ast }$ is the
tortoise coordinate with the following explicit form
\begin{eqnarray}
r_{\ast } &=&\int \frac{dr}{\sqrt{f(r)g(r)}}=r-\frac{A_{0}^{2}}{r_{+}r_{-}}%
\left( \frac{1}{r}-\frac{r_{+}+r_{-}}{r_{+}r_{-}}\ln \left( r\right) \right) 
\nonumber \\
&&+\frac{1}{\left( r_{+}-r_{-}\right) }\left( \frac{A_{0}^{2}+r_{+}^{4}}{%
r_{+}^{2}}\ln \left( r-r_{+}\right) \right.   \nonumber \\
&&\left. -\frac{A_{0}^{2}+r_{-}^{4}}{r_{-}^{2}}\ln \left( r-r_{-}\right)
\right) ,  \label{tortoise}
\end{eqnarray}
that ranges from $-\infty $\ at the event horizon to $+\infty $%
\ at spatial infinity, and note that $r$ in the right-hand side of (\ref{EP}%
) is a function of $r_{\ast }$ by (\ref{tortoise}). Figure \ref{Pot} shows
the behavior of the effective potential (\ref{EP}) versus the tortoise
coordinate for different values of the LQBH parameters $P$ and $A_{0}$. From
this figure, we find that the effect of $P$ on the effective potential is
much more than the minimum area gap $A_{0}$, and therefore it plays a more
important role in the context of LQBH oscillations.

The spectrum of QN modes is a solution to the wave equation (\ref{Weq}) and
we should impose some physically motivated boundary conditions at the
boundaries to find the solutions. The quasinormal boundary conditions imply
that the wave at the event horizon is purely incoming and it is purely
outgoing at spatial infinity, such that
\begin{equation}
\left\{
\begin{array}{c}
\Psi _{l}\left( r_{\ast }\right) \sim e^{-i\omega r_{\ast }}\ \ \ as\ \ \
r_{\ast }\rightarrow -\infty \ (r\rightarrow r_{+}), \\
\Psi _{l}\left( r_{\ast }\right) \sim e^{i\omega r_{\ast }}\ \ \ \ as\ \ \ \
\ \ r_{\ast }\rightarrow \infty \ (r\rightarrow \infty ),%
\end{array}%
\right. .  \label{bc}
\end{equation}

These boundary conditions lead to a discrete set of eigenvalues $\omega
_{nl}=\omega _{r}-i\omega _{i}$\ with a real part giving the actual
oscillation and an imaginary part representing the damping of the
perturbation. The indices of $\omega _{nl}$\ denote the overtone number $n$
and multipole number $l$. In this paper, we investigate the QN modes of
LQBHs by using a couple of independent computational methods, such as the
higher-order Wentzel--Kramers--Brillouin (WKB) approximation and related Pad%
\'{e} approximants, the improved AIM, and the time-domain integration that
we briefly explain in the following subsections.

\subsection{WKB approximation}

The WKB approximation is based on the matching of WKB expansion of the modes
$\Psi _{l}\left( r_{\ast }\right) $ at the event horizon and spatial
infinity with the Taylor expansion near the peak of the potential barrier
through two closely spaced turning points characterized by $\omega
^{2}-V_{l}\left( r_{\ast }\right) =0$. Therefore, the WKB method can be used
for an effective potential that forms a potential barrier and takes zero
values (or small values compared with the height of the barrier) at the
event horizon ($r_{\ast }\rightarrow -\infty $) and spatial infinity ($%
r_{\ast }\rightarrow +\infty $).

This method first applied to the problem of scattering around BHs \cite%
{Schutz}, and is extended to the $3$rd \cite{IyerWill}, $6$th order \cite%
{Konoplya6th}, and $13$th order \cite{Matyjasek13th}. The $13$th order of
WKB approximation is given by the following formula
\begin{eqnarray}
&&\omega ^{2}=V_{0}+\sum_{j=1}^{6}\Omega _{2j}-i\sqrt{-2V_{0}^{\prime \prime
}}\left( n+\frac{1}{2}\right) \times  \notag \\
&&\left( 1+\sum_{j=1}^{6}\Omega _{2j+1}\right) ;\ \ \ \ n=0,1,2,...,
\label{wkb}
\end{eqnarray}%
where $V_{0}$\ denotes the height of the effective potential, $\Omega _{j}$%
's are the WKB correction terms of the $j$th order that depend on the value
of the effective potential and its derivatives at the local maximum, and $n$%
\ is the overtone number.

It is worthwhile to mention that the WKB formula does not give reliable
frequencies for $n\geq l$, while it leads to accurate values for $n<l$\ \
and exact modes in the eikonal limit $l\rightarrow \infty $. We use this
formula up to the $13$th order to calculate the QN frequencies of
perturbations.

On the other hand, one can use Pad\'{e} approximants for the WKB formula (%
\ref{wkb}) to increase the accuracy of this method \cite{Matyjasek13th}. In
order to incorporate the Pad\'{e} approximants, we first define a polynomial
$\mathcal{P}_{k}\left( \varepsilon \right) $ by multiplying the powers of
the order parameter $\varepsilon $ in the WKB correction terms as below \cite%
{KonoplyaPade}%
\begin{eqnarray}
\mathcal{P}_{k}\left( \varepsilon \right) &=&V_{0}+\sum_{j=1}^{6}\varepsilon
^{2j}\Omega _{2j}-i\sqrt{-2V_{0}^{\prime \prime }}\left( n+\frac{1}{2}%
\right) \times  \notag \\
&&\left( \varepsilon +\sum_{j=1}^{6}\varepsilon ^{2j+1}\Omega _{2j+1}\right)
,
\end{eqnarray}%
such that the polynomial order $k$ coincides with the WKB order and the
squared frequency can be obtained by setting $\varepsilon =1$\ as $\omega
^{2}=\mathcal{P}_{k}\left( 1\right) $. Then, we introduce a class of the Pad%
\'{e} approximants $\mathcal{P}_{\tilde{n}/\tilde{m}}\left( \varepsilon
\right) $\ of the polynomial $\mathcal{P}_{k}\left( \varepsilon \right) $
near $\varepsilon =0$ with the condition $k=\tilde{n}+\tilde{m}$\ to obtain
\begin{equation}
\mathcal{P}_{\tilde{n}/\tilde{m}}\left( \varepsilon \right) =\left(
\sum\limits_{i=0}^{\tilde{n}}Q_{i}\varepsilon ^{i}\right) \left/ \left(
\sum\limits_{i=0}^{\tilde{m}}R_{i}\varepsilon ^{i}\right) \right. ,
\label{pade}
\end{equation}%
with%
\begin{equation}
\mathcal{P}_{\tilde{n}/\tilde{m}}\left( \varepsilon \right) -\mathcal{P}%
_{k}\left( \varepsilon \right) =\mathcal{O}\left( \varepsilon ^{k+1}\right) .
\end{equation}

As the next step, since the right-hand side of the WKB formula (\ref{wkb})
is known, we can calculate the coefficients $Q_{i}$'s and $R_{i}$'s of (\ref%
{pade})\ numerically and employ the rational function $\mathcal{P}_{\tilde{n}%
/\tilde{m}}\left( \varepsilon \right) $ to approximate the squared frequency
as $\omega ^{2}=\mathcal{P}_{\tilde{n}/\tilde{m}}\left( 1\right) $.

In most cases, the Pad\'{e} approximation (\ref{pade}) of the order $k=%
\tilde{n}+\tilde{m}$ gives more accurate results for $\tilde{n}\approx
\tilde{m}$ compared to the ordinary WKB formula (\ref{wkb})\ of the same
order \cite{KonoplyaPade}. However, there is no way to choose the
appropriate orders $\tilde{n}$ and $\tilde{m}$ to obtain the frequency with
the highest accuracy. In order to find the suitable orders $\tilde{n}$ and $%
\tilde{m}$, we follow an approach based on averaging of Pad\'{e}
approximations suggested in \cite{KonoplyaPade} so that the
minimum of the \textit{standard deviation (SD) formula} supposed
to specify the most accurate modes.

\subsection{Asymptotic iteration method}

The AIM has been employed to solve the eigenvalue problems and second-order
differential equations \cite{Ciftci,CiftciHall}, and then it was indicated
that an improved version of AIM is an accurate technique for calculating QN modes \cite%
{Naylor,AIM,MomenniaPRD}.

Here, we consider the effect of the LQG correction parameters $P$\ and $%
A_{0} $\ separately to investigate the contribution of either parameter on
the QNMs spectrum and find deviations from those of the Schwarzschild BHs.
Thus, we study the QNMs for different values of one LQG correction parameter
while setting the other one equals to zero. To do so, consider two cases as
follows; one is the $P=0$\ case\ that leads to a single-horizon LQBH with $%
r_{+}=2M$, and the second case is given by $A_{0}=0 $\ which represents LQBHs
with two distinct horizons.

\subsubsection{$P=0$ case}

First, note that the wave equation (\ref{Weq}) has the following form in the
$r$-coordinate
\begin{equation}
f^{2}(r)\Psi ^{\prime \prime }(r)+f(r)f^{\prime }(r)\Psi ^{\prime }(r)+\left[
\omega ^{2}-V(r)\right] \Psi (r)=0,  \label{rweq}
\end{equation}%
where prime denotes the derivative with respect to $r$ and we used the fact
that $g(r)=f(r)$. Equation (\ref{rweq}) is a second-order ordinary differential equation with two regular singular points located at $r=0$ and $r=r_{+}$. In order to apply the boundary conditions (\ref{bc}) to
this differential equation, we follow Leaver \cite{LeaverSchw} and define the following solution%
\begin{eqnarray}
\Psi (r) &=&e^{i\omega \left( r-r_{+}\right) }\left( \frac{r}{r_{+}}\right)
^{i\omega r_{+}}{}\times  \notag \\
&&\left( \frac{r-r_{+}}{r}\right) {}^{-i\omega \left(
A_{0}^{2}+r_{+}^{4}\right) /r_{+}^{3}}\psi \left( r\right) ,  \label{shs}
\end{eqnarray}%
which has the correct asymptotic behavior at the boundaries and $\psi \left(
r\right) $\ is a finite and convergent function. Since the AIM works better
on a compact domain, we also define a new variable $\xi =1-r_{+}/r$. Thus, $%
\xi $\ ranges $0\leq \xi <1$\ so that $\xi \approx 1$\ represents the
spatial infinity and $\xi _{+}=0$ corresponds to the event horizon.

Now, by considering the new variable $\xi $\ and the solution (\ref{shs}),
we can find the standard AIM form of the wave equation (\ref{rweq}) as
follows%
\begin{equation}
\psi ^{\prime \prime }(\xi )=\lambda _{0}\left( \xi \right) \psi ^{\prime
}(\xi )+s_{0}\left( \xi \right) \psi (\xi ),  \label{standardweshbh}
\end{equation}%
where prime denotes the derivative with respect to $\xi $, and $\lambda
_{0}\left( \xi \right) $\ and $s_{0}\left( \xi \right) $\ are
\begin{widetext}
\begin{eqnarray}
\lambda _{0}\left( \xi \right) &=&\frac{2}{1-\xi }-\frac{A_{0}^{2}(1-\xi
)^{3}(3\xi +1)+r_{+}^{4}}{\xi z}+\frac{2i\omega y}{\xi r_{+}^{3}(1-\xi )^{2}}%
,  \label{l0} \\
s_{0}\left( \xi \right) &=&\frac{1}{(1-\xi )^{4}\xi ^{2}r_{+}^{4}}\left\{
\frac{i\omega r_{+}(1-\xi )^{2}\left[ r_{+}^{4}+A_{0}^{2}(1-\xi )^{3}(3\xi
+1)\right] y}{z}\right.  \notag \\
&&+\frac{\xi r_{+}^{4}(1-\xi )^{2}}{z^{2}}\left( l(l+1)z^{2}+r_{+}^{8}(1-\xi
)+A_{0}^{2}(1-\xi )^{8}\left[ \frac{10\xi r_{+}^{4}}{(1-\xi )^{4}}%
-A_{0}^{2}(1+\xi )\right] \right)  \notag \\
&&\left. +\frac{\omega ^{2}}{r_{+}^{2}}\left( y^{2}-z^{2}\right) -2i\omega
\xi r_{+}(1-\xi )y-i\omega r_{+}\left( z+\xi r_{+}^{4}\left[ 2\xi ^{2}(\xi
-4)+9\xi -4\right] \right) \right\} ,  \label{s0}
\end{eqnarray}
\end{widetext}with $y=2\xi r_{+}^{4}(\xi -2)+r_{+}^{4}+A_{0}^{2}(1-\xi )^{2}$
and $z=r_{+}^{4}+A_{0}^{2}(1-\xi )^{4}$.

\subsubsection{$A_{0}=0$ case}

On the other hand, as for the $P=0$\ case, we also obtain the standard AIM
form of the wave equation for $A_{0}=0$. The wave equation (\ref{Weq}) has
the following form in the $r$-coordinate
\begin{eqnarray}
&&f(r)g(r)\Psi ^{\prime \prime }(r)+\partial _{r}\left[ \sqrt{f(r)g(r)}%
\right] \sqrt{f(r)g(r)}\Psi ^{\prime }(r)  \notag \\
&&+\left[ \omega ^{2}-V(r)\right] \Psi (r)=0.  \label{Rweq}
\end{eqnarray}

In this case, we deal with BHs with two horizons located at $r_{-}$\ and $r_{+}$, and therefore the differential equation (\ref{Rweq}) contains three regular singular points located at $r=0$, $r=r_{-}$, and $r=r_{+}$. Following \cite{LeaverRN}, we define the solution
\begin{eqnarray}
\Psi (r) &=&e^{i\omega r}r^{-1}\left( r-r_{-}\right) {}^{1+i\omega r_{+} +i\omega
r_{+}^{2}/\left( r_{+}-r_{-}\right) }\times  \notag \\
&&\left( r-r_{+}\right) {}^{-i\omega r_{+}^{2}/\left( r_{+}-r_{-}\right)
}\psi (r),  \label{dhs}
\end{eqnarray}%
to apply the boundary conditions (\ref{bc}) such that $\psi \left(
r\right) $\ is a finite and convergent function. One may note that the solutions (\ref{shs}) and (\ref{dhs}) are not consistent in the common limit $A_{0}=0=P$. In this regard, we should mention that the constant $e^{-i\omega r_{+}} r_{+}^{-i\omega r_{+}}$ was multiplied to the solution (\ref{shs}) by hand to obtain a simpler form for the relations (\ref{l0}) and (\ref{s0}).

Now, we can find the
standard AIM form of the wave equation (\ref{Rweq}) with the help of the new
variable $\xi $\ and the solution (\ref{dhs}) as below%
\begin{equation}
\psi ^{\prime \prime }(\xi )=\hat{\lambda}_{0}\left( \xi \right) \psi
^{\prime }(\xi )+\hat{s}_{0}\left( \xi \right) \psi (\xi ),
\label{standardwedhbh}
\end{equation}%
where $\hat{\lambda}_{0}\left( \xi \right) $\ and $\hat{s}_{0}\left( \xi
\right) $\ are
\begin{widetext}
\begin{eqnarray}
\hat{\lambda}_{0}\left( \xi \right)  &=&\frac{r_{+}\left[ 2i\omega
r_{+}\left( 2\xi ^{2}-4\xi +1\right) -3\xi ^{2}+4\xi -1\right] +r_{-}(1-\xi )%
\left[ 1+\xi \left( 2i\omega r_{+}+6\xi -7\right) \right] }{\xi (1-\xi )^{2}%
\left[ r_{+}-r_{-}(1-\xi )\right] }, \\
\hat{s}_{0}\left( \xi \right)  &=&\frac{1}{\xi ^{2}(1-\xi )^{4}\left[
r_{+}-(1-\xi )r_{-}\right] ^{2}}\left[ r_{+}^{4}V(\xi )-\xi (1-\xi )\times
\right.   \nonumber \\
&&\left\{ r_{-}(1-\xi )^{2}\left[ r_{+}(1-3\xi )-r_{-}(1-\xi )(1-4\xi )%
\right] \right.  \nonumber\\
&&+i\omega r_{+}(1-\xi )\left[ 4r_{+}^{2}(1-\xi )-r_{+}r_{-}\left( 8\xi
^{2}-18\xi +7\right) +r_{-}^{2}(1-\xi )(1-4\xi )\right]   \nonumber \\
&&\left. \left. -r_{+}^{2}\omega ^{2}\left[ r_{-}\left( 1-\xi \right)
-2r_{+}(2-\xi )\right] \left[ r_{-}\xi +2r_{+}(1-\xi )\right] \right\} %
\right] .
\end{eqnarray}
\end{widetext}

Once the standard AIM form of the master wave equation is obtained in (\ref%
{standardweshbh})\ and (\ref{standardwedhbh}), we can express higher
derivatives of $\psi \left( \xi \right) $ in terms of $\psi \left( \xi
\right) $ and $\psi ^{\prime }\left( \xi \right) $ as follows%
\begin{equation}
\psi ^{(n+2)}\left( \xi \right) =\lambda _{n}\left( \xi \right) \psi
^{\prime }\left( \xi \right) +s_{n}\left( \xi \right) \psi \left( \xi
\right) ,
\end{equation}%
with the recurrence relations%
\begin{equation}
\begin{array}{c}
\lambda _{n}\left( \xi \right) =\lambda _{n-1}^{\prime }\left( \xi \right)
+s_{n-1}\left( \xi \right) +\lambda _{0}\left( \xi \right) \lambda
_{n-1}\left( \xi \right) , \\
\\
s_{n}\left( \xi \right) =s_{n-1}^{\prime }\left( \xi \right) +s_{0}\left(
\xi \right) \lambda _{n-1}\left( \xi \right) .%
\end{array}
\label{RR}
\end{equation}

We now expand $\lambda _{n}\left( \xi \right) $ and $s_{n}\left( \xi \right)
$ in a Taylor series around some point $\bar{\xi}$ at which the AIM is
performed%
\begin{equation}
\begin{array}{c}
\lambda _{n}\left( \xi \right) =\sum\limits_{j=0}^{\infty }c_{n}^{j}\left(
\xi -\bar{\xi}\right) ^{j}, \\
\\
s_{n}\left( \xi \right) =\sum\limits_{j=0}^{\infty }d_{n}^{j}\left( \xi -%
\bar{\xi}\right) ^{j},%
\end{array}%
\end{equation}%
which allows us to rewrite the recurrence relations (\ref{RR}) in terms of
the series coefficients $c_{n}^{j}$ and $d_{n}^{j}$%
\begin{equation}
c_{n}^{j}=\left( j+1\right)
c_{n-1}^{j+1}+d_{n-1}^{j}+\sum\limits_{k=0}^{j}c_{0}^{k}c_{n-1}^{j-k},
\end{equation}%
\begin{equation}
d_{n}^{j}=\left( j+1\right)
d_{n-1}^{j+1}+\sum\limits_{k=0}^{j}d_{0}^{k}c_{n-1}^{j-k}.
\end{equation}

For sufficiently large $n$, we consider the following termination to the
number of iterations%
\begin{equation}
\frac{s_{n}\left( \xi \right) }{\lambda _{n}\left( \xi \right) }=\frac{%
s_{n-1}\left( \xi \right) }{\lambda _{n-1}\left( \xi \right) },
\end{equation}%
which leads to%
\begin{equation}
\delta _{n}=s_{n}\left( \xi \right) \lambda _{n-1}\left( \xi \right)
-s_{n-1}\left( \xi \right) \lambda _{n}\left( \xi \right) =0,
\end{equation}%
and in terms of the Taylor series coefficients, we have%
\begin{equation}
d_{n}^{0}c_{n-1}^{0}-d_{n-1}^{0}c_{n}^{0}=0,  \label{QC}
\end{equation}%
that is known as the \textit{quantization condition }and gives an equation
in terms of the QN frequencies $\omega $. As the next step, we fix all the
free parameters, namely, the multipole number $l$, the BH mass $M$, the
polymeric function $P$, and the minimum area gap of LQG $A_{0}$. Finally, we
use the \textit{quantization condition}\ (\ref{QC})\ and a root finder to
calculate the QN modes.

\subsection{Ringdown waveform}

In order to investigate the contribution of all modes, we can integrate the
wave-like equation (\ref{Weq}) on a finite time domain. This also helps us
to explore the time evolution of modes and dynamical stability of the BH
case study. To do so, we follow \cite{Gundlach}\ and write the perturbation
equation (\ref{Weq}) in terms of the light-cone coordinates $u=t-r_{\ast }$\
and $v=t+r_{\ast }$ in the following form%
\begin{equation}
-4\frac{\partial ^{2}\Psi _{l}\left( u,v\right) }{\partial u\partial v}%
=V_{l}\left( u,v\right) \Psi _{l}\left( u,v\right) ,  \label{uveq}
\end{equation}%
where $\Psi _{l}$\ assumed to have time dependence $e^{-i\omega t}$. To find
a unique solution to (\ref{uveq}), the initial data must be specified on the
two null surfaces $u=u_{0}$\ and $v=v_{0}$. Here, we set $\Psi _{l}\left(
u,0\right) =1$\ at $v=0$,\ and use the Gaussian wave packet%
\begin{equation}
\Psi _{l}\left( 0,v\right) =\exp \left( -\frac{\left( v-v_{c}\right) ^{2}}{%
2\sigma ^{2}}\right) ,
\end{equation}%
centered on $v_{c}$ and having width $\sigma $ at $u=0$. Then, we
choose the observer to be located at $r=5r_{+}$ and use built-in
Wolfram Mathematica commands for solving partial differential
equations to generate the ringdown waveforms.\ Finally, we employ
the Prony method \cite{Marple,Prony}, a method for mining
information from (damped) sinusoidal signals, to extract dominant
(fundamental) frequency from the data generated in the ringdown
profile.

\section{Quasinormal Modes \label{QNMs}}

Before investigating the QN oscillations by using the mentioned methods in
general, let us first reconstruct Tables I and II of \cite{LQBHJamil}\
by employing the $6$th order WKB approximation. We present our results in
Tables \ref{6thorderWKBscalar} and \ref{6thorderWKBelectromagnetic} with the
relative error $\left\vert \left( \mathring{\omega}-\omega \right) /\omega
\right\vert \times 100\%$, where $\mathring{\omega}$'s are given in \cite%
{LQBHJamil}\ through Tables I and II, and $\omega $'s\ are presented in
our Tables \ref{6thorderWKBscalar} and \ref{6thorderWKBelectromagnetic}.
Although both $\mathring{\omega}$\ and $\omega $\ were calculated by
employing the $6$th order WKB formula, our tables indicate a disagreement
between $\mathring{\omega}$\ and $\omega $. The error increases as the
polymeric function increases and it is about $30\%$ in the worst case. We
found that the WKB method was not properly used which led to this error (see
\cite{KonoplyaPade} to find popular mistakes when employing the WKB
approximation). Therefore, figures $9$ and $10$ illustrated in \cite%
{LQBHJamil}\ should be modified according to the following tables as well.

\begin{center}
\begin{table*}[tbh]
\scalebox{.9} {\begin{tabular}{|c|c|c|c|c|} \hline\hline $P$ & &
$\omega _{01}$ & $\omega _{02}$ & $\omega _{12}$ \\ \hline\hline
$0.0$ &  & $0.2929-0.0978i(0.00\%)$ & $0.4836-0.0968i(0.00\%)$ & $0.4638-0.2956i(0.00\%)$ \\ \hline
$0.1$ &  & $0.3739-0.1192i(4.62\%)$ & $0.6206-0.1184i(2.71\%)$ & $0.5987-0.3612i(7.82\%)$ \\ \hline
$0.2$ &  & $0.4652-0.1405i(9.07\%)$ & $0.7757-0.1400i(5.35\%)$ & $0.7528-0.4260i(15.6\%)$ \\ \hline
$0.3$ &  & $0.5653-0.1603i(13.1\%)$ & $0.9463-0.1600i(7.82\%)$ & $0.9236-0.4860i(23.1\%)$ \\ \hline
$0.4$ &  & $0.6717-0.1770i(16.7\%)$ & $1.1280-0.1771i(9.99\%)$ & $1.1067-0.5366i(29.8\%)$ \\ \hline\hline
\end{tabular}}
\caption{The QNMs $\protect\omega _{nl}$ of scalar perturbations for $M=1$
and $A_{0}=0.01$ calculated by the $6$th order WKB formula.}
\label{6thorderWKBscalar}
\end{table*}
\end{center}


\begin{center}
\begin{table*}[tbh]
\scalebox{.9} {\begin{tabular}{|c|c|c|c|c|} \hline\hline $P$ & &
$\omega _{01}$ & $\omega _{02}$ & $\omega _{12}$ \\ \hline\hline
$0.0$ &  & $0.2482-0.0926i(0.00\%)$ & $0.4576-0.0950i(0.00\%)$ & $0.4365-0.2907i(0.00\%)$ \\ \hline
$0.1$ &  & $0.3226-0.1141i(4.90\%)$ & $0.5908-0.1167i(2.77\%)$ & $0.5676-0.3562i(8.03\%)$ \\ \hline
$0.2$ &  & $0.4084-0.1357i(9.58\%)$ & $0.7426-0.1383i(5.48\%)$ & $0.7184-0.4214i(16.0\%)$ \\ \hline
$0.3$ &  & $0.5043-0.1560i(13.9\%)$ & $0.9108-0.1586i(7.99\%)$ & $0.8869-0.4820i(23.7\%)$ \\ \hline
$0.4$ &  & $0.6082-0.1734i(17.8\%)$ & $1.0911-0.1759i(10.2\%)$ & $1.0686-0.5334i(30.5\%)$ \\ \hline\hline
\end{tabular}}
\caption{The QNMs $\protect\omega _{nl}$ of electromagnetic perturbations
for $M=1 $ and $A_{0}=0.01$ calculated by the $6$th order WKB formula.}
\label{6thorderWKBelectromagnetic}
\end{table*}
\end{center}


\begin{center}
\begin{table}[tbh]
\scalebox{0.9} {\begin{tabular}{|c|c|c|c|c|} \hline\hline $A_{0}$
&  & $\omega _{00}$ & $\omega _{01}$ & $\omega _{02}$ \\
\hline\hline $0$ &  & $\begin{array}{c}
0.2209-0.2098i \\
-\end{array}$ & $\begin{array}{c}
0.5859-0.1953i \\
0.5858-0.1955i\end{array}$ & $\begin{array}{c}
0.9673-0.1935i \\
0.9673-0.1935i\end{array}$ \\ \hline $0.25$ &  & $\begin{array}{c}
0.2155-0.2083i \\
-\end{array}$ & $\begin{array}{c}
0.5779-0.1945i \\
0.5777-0.1952i\end{array}$ & $\begin{array}{c}
0.9552-0.1928i \\
0.9553-0.1927i\end{array}$ \\ \hline $0.5$ &  & $\begin{array}{c}
0.2008-0.2057i \\
-\end{array}$ & $\begin{array}{c}
0.5576-0.1930i \\
0.5537-0.1977i\end{array}$ & $\begin{array}{c}
0.9245-0.1913i \\
0.9246-0.1912i\end{array}$ \\ \hline\hline
\end{tabular}}
\caption{The fundamental QNM ($n=0$) for $P=0$\ and different values of $%
A_{0}$ and $l$ calculated by the AIM (first row) and the $6$th order WKB
formula (second row).}
\label{tab3}
\end{table}
\end{center}


\begin{center}
\begin{table}[tbh]
\scalebox{0.9} {\begin{tabular}{|c|c|c|c|c|} \hline\hline $A_{0}$
&  & $\omega _{11}$ & $\omega _{12}$ & $\omega _{13}$ \\
\hline\hline $0$ &  & $\begin{array}{c}
0.5289-0.6125i \\
-\end{array}$ & $\begin{array}{c}
0.9277-0.5912i \\
0.9277-0.5913i\end{array}$ & $\begin{array}{c}
1.3213-0.5846i \\
1.3213-0.5846i\end{array}$ \\ \hline $0.25$ &  & $\begin{array}{c}
0.5130-0.6080i \\
-\end{array}$ & $\begin{array}{c}
0.9114-0.5881i \\
0.9116-0.5878i\end{array}$ & $\begin{array}{c}
1.3020-0.5819i \\
1.3020-0.5818i\end{array}$ \\ \hline $0.5$ &  & $\begin{array}{c}
0.4703-0.6013i \\
-\end{array}$ & $\begin{array}{c}
0.8696-0.5825i \\
0.8694-0.5836i\end{array}$ & $\begin{array}{c}
1.2527-0.5768i \\
1.2528-0.5767i\end{array}$ \\ \hline\hline
\end{tabular}}
\caption{The first overtone ($n=1$) for $P=0$\ and different values of $%
A_{0} $ and $l$ calculated by the AIM (first row) and the $6$th order WKB
formula (second row).}
\label{tab4}
\end{table}
\end{center}


\begin{center}
\begin{table}[tbh]
\scalebox{0.9}
{\begin{tabular}{|c|c|c|c|}
\hline\hline
$A_{0}$ &  & $\omega _{22}$ & $\omega _{23}$ \\ \hline\hline
$0$ &  & $%
\begin{array}{c}
0.8611-1.0171i \\
-%
\end{array}%
$ & $%
\begin{array}{c}
1.2673-0.9920i \\
1.2672-0.9920i%
\end{array}%
$ \\ \hline
$0.25$ &  & $%
\begin{array}{c}
0.8347-1.010i \\
-%
\end{array}%
$ & $%
\begin{array}{c}
1.2414-0.9863i \\
1.2414-0.9860i%
\end{array}%
$ \\ \hline
$0.5$ &  & $%
\begin{array}{c}
0.7653-0.9985i \\
-%
\end{array}%
$ & $%
\begin{array}{c}
1.1754-0.9761i \\
1.1755-0.9762i%
\end{array}%
$ \\ \hline\hline
\end{tabular}}%
\caption{The second overtone ($n=2$) for $P=0$\ and different values of $%
A_{0}$ and $l$ calculated by the AIM (first row) and the $6$th order WKB
formula (second row).}
\label{tab5}
\end{table}
\end{center}


\begin{center}
\begin{table}[tbh]
\scalebox{0.9} {\begin{tabular}{|c|c|c|c|c|} \hline\hline $P$ & &
$\omega _{00}$ & $\omega _{01}$ & $\omega _{02}$ \\ \hline\hline
$0$ &  & $\begin{array}{c}
0.2209-0.2098i \\
-\end{array}$ & $\begin{array}{c}
0.5859-0.1953i \\
0.5858-0.1955i\end{array}$ & $\begin{array}{c}
0.9673-0.1935i \\
0.9673-0.1935i\end{array}$ \\ \hline $0.01$ &  & $\begin{array}{c}
0.2253-0.2140i \\
-\end{array}$ & $\begin{array}{c}
0.6011-0.1996i \\
0.6010-0.1998i\end{array}$ & $\begin{array}{c}
0.9930-0.1978i \\
0.9930-0.1978i\end{array}$ \\ \hline $0.02$ &  & $\begin{array}{c}
0.2298-0.2182i \\
-\end{array}$ & $\begin{array}{c}
0.6165-0.2038i \\
0.6165-0.2040i\end{array}$ & $\begin{array}{c}
1.0190-0.2021i \\
1.0190-0.2021i\end{array}$ \\ \hline\hline
\end{tabular}}
\caption{The fundamental QNM ($n=0$) for $A_{0}=0$\ and different values of $%
P$ and $l$ calculated by the AIM (first row) and the $6$th order WKB formula
(second row).}
\label{tab6}
\end{table}
\end{center}


\begin{center}
\begin{table}[tbh]
\scalebox{0.9} {\begin{tabular}{|c|c|c|c|c|} \hline\hline $P$ & &
$\omega _{11}$ & $\omega _{12}$ & $\omega _{13}$ \\ \hline\hline
$0$ &  & $\begin{array}{c}
0.5289-0.6125i \\
-\end{array}$ & $\begin{array}{c}
0.9277-0.5912i \\
0.9277-0.5913i\end{array}$ & $\begin{array}{c}
1.3213-0.5846i \\
1.3213-0.5846i\end{array}$ \\ \hline $0.01$ &  & $\begin{array}{c}
0.5433-0.6256i \\
-\end{array}$ & $\begin{array}{c}
0.9529-0.6042i \\
0.9529-0.6042i\end{array}$ & $\begin{array}{c}
1.3571-0.5975i \\
1.3571-0.5975i\end{array}$ \\ \hline $0.02$ &  & $\begin{array}{c}
0.5580-0.6387i \\
-\end{array}$ & $\begin{array}{c}
0.9785-0.6172i \\
0.9785-0.6172i\end{array}$ & $\begin{array}{c}
1.3933-0.6105i \\
1.3933-0.6105i\end{array}$ \\ \hline\hline
\end{tabular}}
\caption{The first overtone ($n=1$) for $A_{0}=0$\ and different values of $%
P $ and $l$ calculated by the AIM (first row) and the $6$th order WKB
formula (second row).}
\label{tab7}
\end{table}
\end{center}


\begin{center}
\begin{table}[tbh]
\scalebox{0.9}
{\begin{tabular}{|c|c|c|c|}
\hline\hline
$P$ &  & $\omega _{22}$ & $\omega _{23}$ \\ \hline\hline
$0$ &  & $%
\begin{array}{c}
0.8611-1.0171i \\
-%
\end{array}%
$ & $%
\begin{array}{c}
1.2673-0.9920i \\
1.2672-0.9920i%
\end{array}%
$ \\ \hline
$0.01$ &  & $%
\begin{array}{c}
0.8854-1.0391i \\
-%
\end{array}%
$ & $%
\begin{array}{c}
1.3023-1.0137i \\
1.3022-1.0138i%
\end{array}%
$ \\ \hline
$0.02$ &  & $%
\begin{array}{c}
0.9102-1.0611i \\
-%
\end{array}%
$ & $%
\begin{array}{c}
1.3379-1.0356i \\
1.3379-1.0356i%
\end{array}%
$ \\ \hline\hline
\end{tabular}}%
\caption{The second overtone ($n=2$) for $A_{0}=0$\ and different values of $%
P$ and $l$ calculated by the AIM (first row) and the $6$th order WKB formula
(second row).}
\label{tab8}
\end{table}
\end{center}


\begin{figure*}[tbh]
\centering
\includegraphics[width=0.35\textwidth]{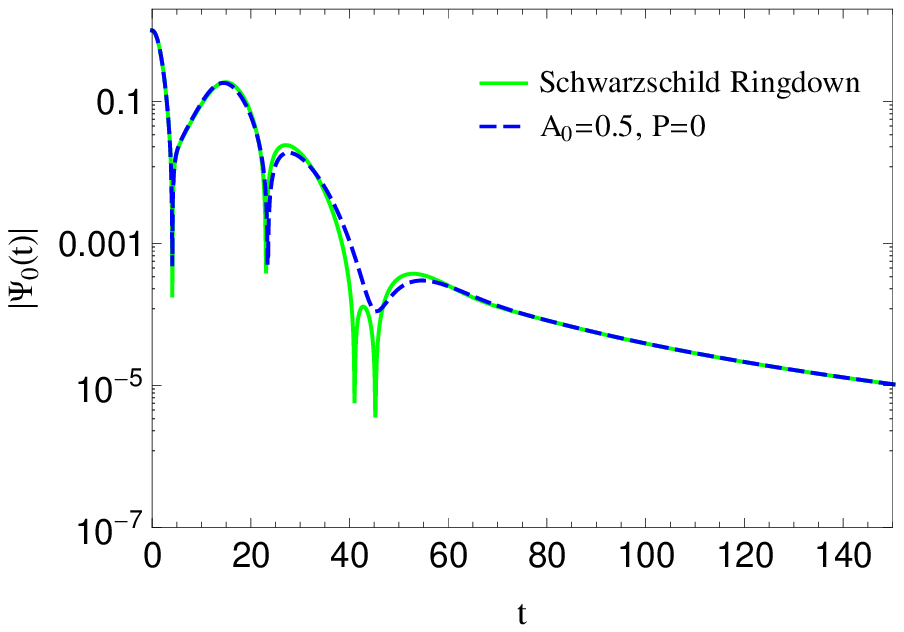} \includegraphics[width=0.35%
\textwidth]{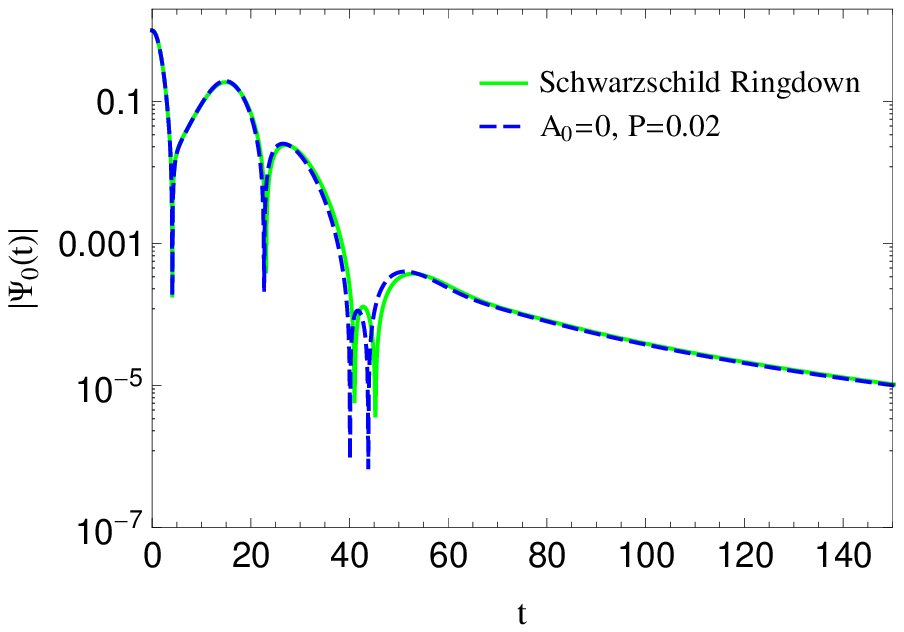}
\caption{This figure evaluated at $r=5r_{+}$ for $l=0$. Either panel
indicates the time evolution of the wave function $\Psi _{0}\left( t\right) $
of scalar perturbations for fixed value of one LQG correction parameter
while setting the other one equals to zero.}
\label{TDPl0}
\end{figure*}
\begin{figure*}[tbh]
\centering
\includegraphics[width=0.35\textwidth]{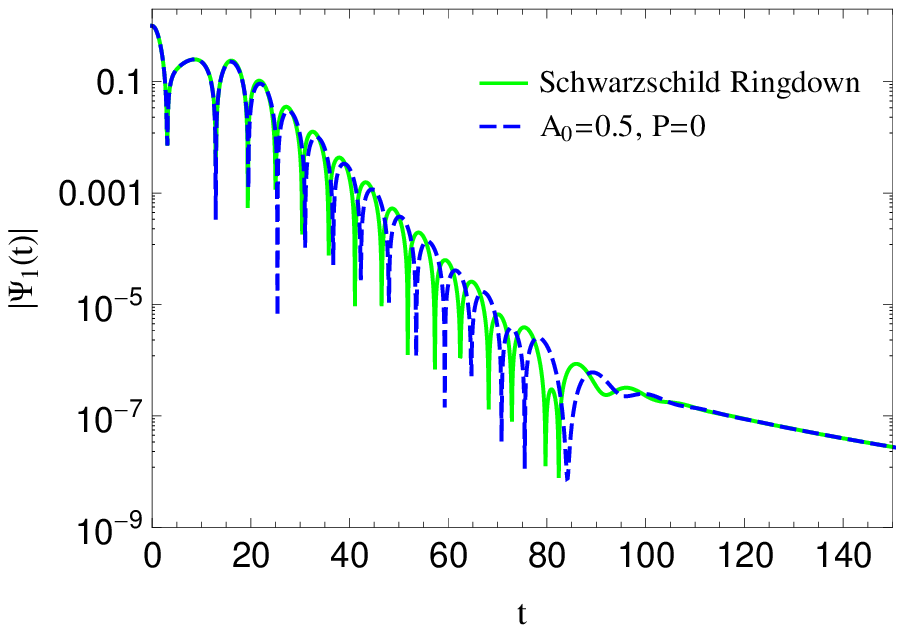} \includegraphics[width=0.35%
\textwidth]{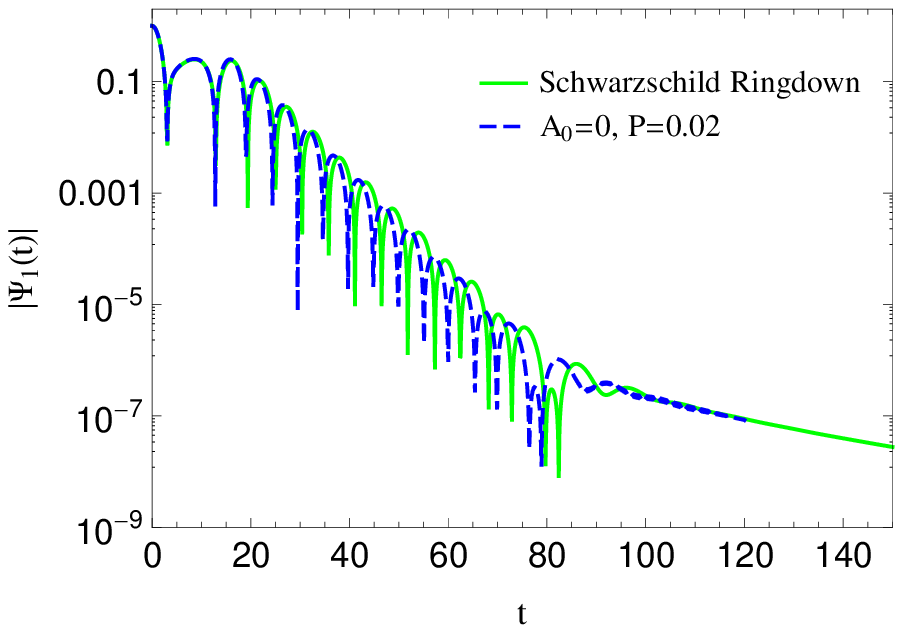}
\caption{This figure evaluated at $r=5r_{+}$ for $l=1$. Either panel
indicates the time evolution of the wave function $\Psi _{1}\left( t\right) $
of scalar perturbations at early, intermediate, and late times. The ringdown
waveform is plotted for fixed value of one LQG correction parameter while
setting the other one equals to zero.}
\label{TDPl1}
\end{figure*}
\begin{figure*}[tbh]
\centering
\includegraphics[width=0.35\textwidth]{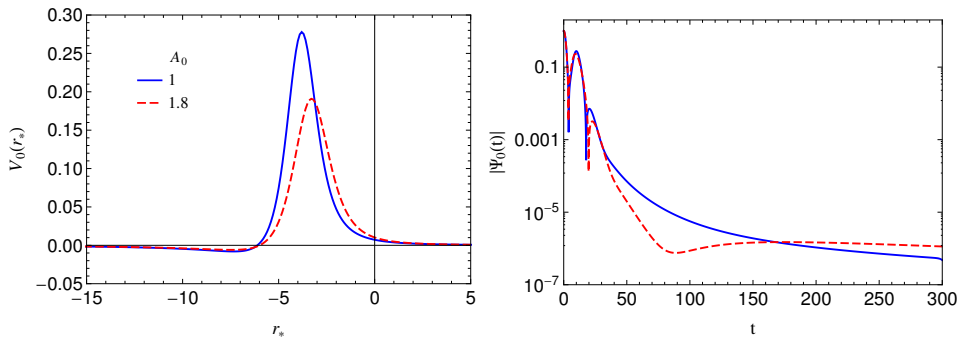} \includegraphics[width=0.35%
\textwidth]{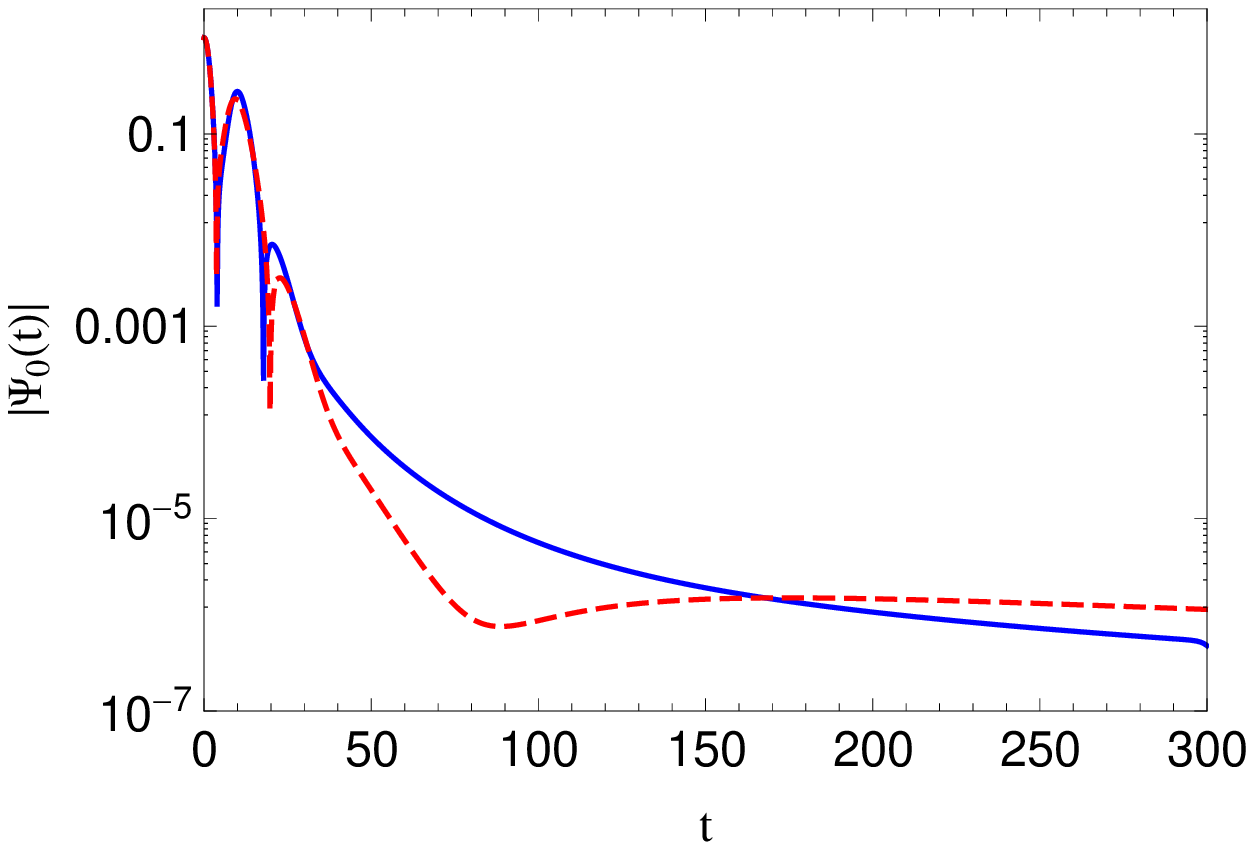}
\caption{The effective potential versus tortoise coordinate for $l=0 $ and $%
P=0.9$ with a negative gap (left panel), and the time evolution of the
corresponding mode $\Psi _{0}\left( t\right) $ (right panel).}
\label{StableA1}
\end{figure*}
\begin{figure*}[tbh]
\centering
\includegraphics[width=0.35\textwidth]{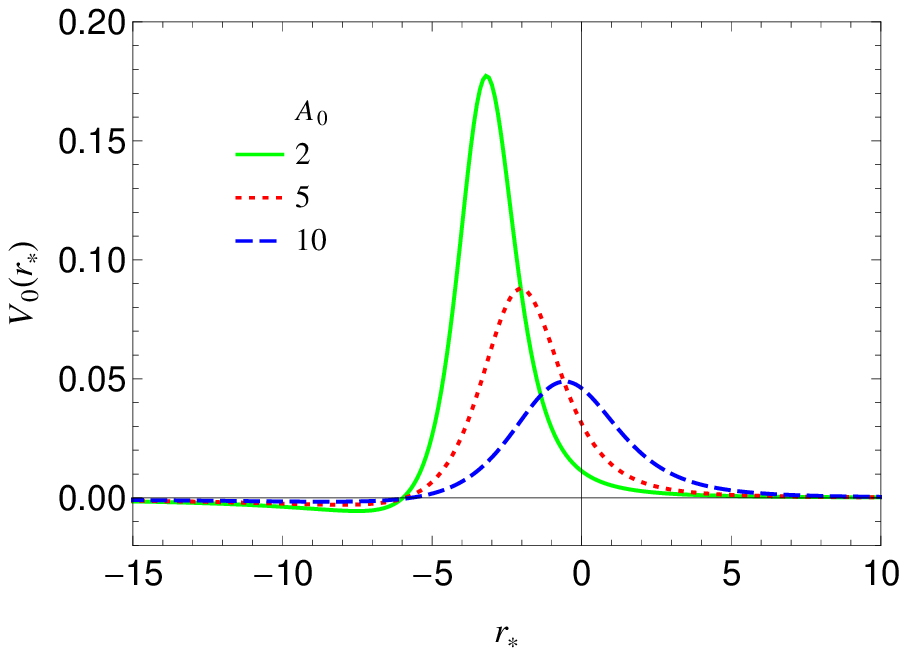} \includegraphics[width=0.35%
\textwidth]{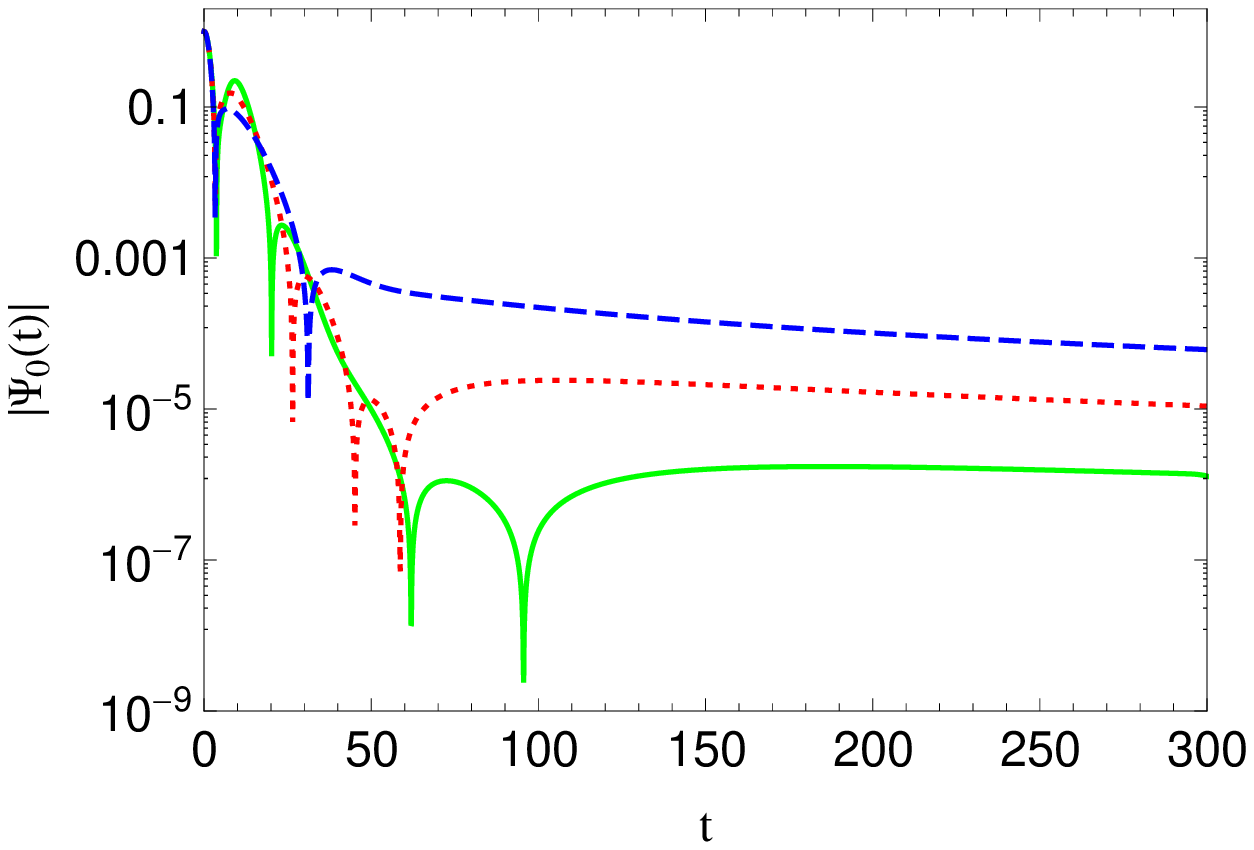}
\caption{The effective potential versus tortoise coordinate for $l=0 $ and $%
P=0.9$ with a negative gap (left panel), and the time evolution of the
corresponding mode $\Psi _{0}\left( t\right) $ (right panel).}
\label{StableA2}
\end{figure*}

Now, we look for the lowest overtone and obtain the QNMs for
various values of the free parameters $P$\ and $A_{0}$\ to
investigate the effects of the LQG corrections on the QN
frequencies and find deviations from those of the Schwarzschild
BHs. Tables \ref{tab3}-\ref{tab5} show the effect of the minimum
area gap of LQG $A_{0}$\ on the QN frequencies. Although the free
parameter $A_{0}$\ is a small quantity, we have chosen large
values to
see its effects on the QN frequencies. Besides, Tables \ref{tab6}-\ref%
{tab8} show the effect of the polymeric function $P$\ on the QN frequencies.
The QNMs were calculated for $M=1/2$ and the rows corresponding to $A_{0}=0$
and $P=0$\ indicate the Schwarzschild QN frequencies.

By considering Tables \ref{tab3}-\ref{tab5}, one can see that both the real
and imaginary parts of the QN frequencies decrease with an increase in $%
A_{0} $. Therefore, the perturbations in the background spacetime
of LQBHs with non-zero $A_{0}$\ live longer with fewer
oscillations in comparison with the Schwarzschild solutions.
However, Tables \ref{tab6}-\ref{tab8}\ show an opposite behavior
for the polymeric function $P$. In this case, the real part of
frequencies and damping rate increase as $P$ increases, and thus,
the perturbations in the background of LQBHs with non-zero $P$\
enjoy faster decay with more oscillations compared to the
Schwarzschild BHs. In Ref. \cite{BarrauUniverse}, it has been
stated that the polymerization does not affect the damping rate of
QNMs, whereas one can obviously see the effects of $P$ on the
imaginary part of the QN frequencies in Tables
\ref{tab6}-\ref{tab8}. However, note that the polymeric function
affects the real part much more than the imaginary part, and this
fact may lead to a misleading conclusion so that the
polymerization does not affect the damping rate. We also see that
the $6$th order WKB formula is in good agreement with the AIM
results for $n<l$ and low values of the LQG correction parameters
(we shall discuss the higher-order WKB formula and related
Pad\'{e} approximants in the next section).

The effects of $A_{0}$\ and $P$ on the QNMs that are described
above, do not exactly coincide with the picture given in
\cite{Chen2011} as well. This is because the dominant fundamental
mode $n=0=l$\ was calculated with the help of usual $3$rd order
WKB formula while this formula does not give reliable
frequencies for $n\geq l$. More importantly, for some higher values of $%
A_{0} $\ and $P$ (say $A_{0}\geq 1$\ and $P\geq 0.9$\ that was considered in
\cite{Chen2011}), a negative gap appears in the effective potential (\ref{EP}%
) such that the WKB expansion could not be performed. This
negative gap appears for the lowest multipole number ($l=0$) that
may lead to instability (see Figs. \ref{StableA1}-\ref{StableA2}
and related discussion below).

From Tables \ref{tab3}-\ref{tab8}, we find two important differences between
the LQG correction quantities $A_{0}$ and $P$. First, one can see that the
effects of the polymeric function $P$\ on the QNMs are much higher than the
minimum area gap $A_{0}$. Therefore, the polymeric function $P$ plays a more
important role in the evolution of fields on the background geometry of
LQBHs compared with $A_{0}$. Second, the LQG correction parameters affect
the value of the QNMs differently. In other words, both the real and
imaginary parts decrease as $A_{0}$\ increases, whereas they increase as $P$%
\ increases.

Furthermore, the time-domain profile of modes is illustrated in Figs. \ref%
{TDPl0}\ and \ref{TDPl1} for fixed value of one LQG correction parameter
while setting the other one equal to zero. According to the time evolution
of modes, we can observe three different stages of QN oscillations of the
wave function $\Psi _{l}\left( t,r\right) $ at early, intermediate, and late
times for $l=0,1$. First, note that by considering the contribution of all
modes, both the real and imaginary parts still decrease as $A_{0}$\
increases (the left panels of Figs. \ref{TDPl0}\ and \ref{TDPl1}) whereas
they increase as $P$\ increases (the right panels of Figs. \ref{TDPl0}\ and %
\ref{TDPl1}) that confirm results deduced from Tables \ref{tab3}-\ref{tab8}.
Second, we see that although the time evolution of scalar field in the
background of LQBHs differ from the Schwarzschild ones at intermediate
times, this is not the case for the late times and both BH solutions seem to
share the same power-law tail as $\Psi _{l}\left( t\right) \sim t^{-(2l+3)}$
\cite{SchwTail}.

In addition, by employing the Prony method to fit the data in Figs. \ref%
{TDPl0}\ and \ref{TDPl1}, we calculated the longest-lived modes
$\omega_{00}$ as
\begin{equation}
\left\{
\begin{array}{cc}
0.221015-0.209788i, & \text{for Schwarzschild BH, } \\
0.200820-0.205698i, & \text{for }A_{0}=0.5\text{, }P=0\text{, } \\
0.229522-0.218754i, & \text{for }A_{0}=0\text{, }P=0.02\text{,}%
\end{array}%
\right.
\end{equation}%
for $l=0$\ which coincide with results in Tables \ref{tab3} and \ref{tab6}.
The results of $\omega _{01}$\ (for $l=1$) are
\begin{equation}
\left\{
\begin{array}{cc}
0.585883-0.195323i, & \text{for Schwarzschild BH, } \\
0.557644-0.192999i, & \text{for }A_{0}=0.5\text{, }P=0\text{, } \\
0.616512-0.203837i, & \text{for }A_{0}=0\text{, }P=0.02\text{,}%
\end{array}%
\right.
\end{equation}%
and they are in a good agreement with the Tables \ref{tab3} and \ref{tab6} as
well.

As for the dynamic stability of our BH case study, Figs. \ref{TDPl0}\ and %
\ref{TDPl1} show that the perturbations decay in time for small values of $%
A_{0}$\ and $P$, and also, the effective potential (\ref{EP})\ is positive
definite (see Fig. \ref{Pot}). These conditions guarantee the dynamical
stability of the LQBHs undergoing scalar perturbations.

We recall that for some higher values of $A_{0}$\ and $P$ (say $A_{0}\geq 1$%
\ and $P\geq 0.9$), a negative gap appears in the effective potential for
the lowest multipole number (see the left panel of Figs. \ref{StableA1}\ and %
\ref{StableA2}). This negative gap may lead to a bound state with negative
energy, hence a growing mode will appear in the spectrum and dominate at
late time which means dynamic instability (see \cite%
{InstabilityKZ,InstabilityWang}\ as examples of dynamic instability of low-$%
l $ modes). Therefore, we should check the stability for this case
numerically while the contribution of all the modes is taken into account.

The right panel of Figs. \ref{StableA1}\ and \ref{StableA2} show that the
perturbations decay in time that indicate the dynamical stability of the
BHs. In the right panel of Fig. \ref{StableA1} and for $A_{0}=1.8$, the
asymptotic tail of modes first starts to grow but finally decays at late
time. However, note that the LQG correction parameters $A_{0}$\ and $P$ are
very small quantities by definition, and we examined the large $A_{0}$\ and $%
P$ case to complete the discussion. The important point is that
the BHs are dynamically stable for small $A_{0}$\ and $P$ as
demonstrated in Figs. \ref{Pot}-\ref{TDPl1}.

\subsection{Higher-order WKB formula and Pad\'{e} approximants}

As the final remark, we should note that, usually, employing the numerical
methods to obtain the QNMs is hard, and normally one needs to modify the
approach based on the different effective potentials. On the other hand, the
WKB approximation provides quite a simple, powerful, and accurate tool for
investigating the dynamical properties of BHs in some cases. However,
generally, this method does not always give a reliable result and neither
guarantees a good estimation for the error \cite{KonoplyaPade}. Besides, we
cannot always increase the WKB order to obtain a more accurate frequency due
to the fact that the WKB formula (\ref{wkb}) asymptotically approaches the
QNMs. So, there is an order of the WKB formula that provides the best
approximation and the error increases as the order of the formula increases.
Thus, it will be helpful to find the most accurate WKB order and related Pad%
\'{e} approximation for calculating the QN frequencies of LQBHs.

In order to estimate the error of the WKB approximation (\ref{wkb}), we use
the following quantity \cite{KonoplyaPade}%
\begin{equation}
\Delta _{k}=\frac{\left\vert \omega _{k+1}-\omega _{k-1}\right\vert }{2},
\end{equation}%
because each WKB correction term affects either the real or imaginary part
of the squared frequencies. This relation obtains the error estimation of $%
\omega _{k}$ that is calculated with the WKB formula of the $k$th\ order,
and the minimum value of $\Delta _{k}$ usually gives the WKB order in which
the error is minimal. It was shown that $\Delta _{k}$\ provides a good
estimation of the error order for the Schwarzschild BH, usually satisfying
\cite{KonoplyaPade}
\begin{equation}
\Delta _{k}\gtrsim \delta _{k}=\left\vert \omega -\omega _{k}\right\vert ,
\label{DKbound}
\end{equation}%
where $\omega $ is the accurate value of the quasinormal frequency. The
quantity $\Delta _{k}$ has been also used to estimate the error of WKB
formula in conformal Weyl gravity \cite{MomenniaWeyl}, and the results
mostly have satisfied the condition (\ref{DKbound}) as well.

Here, we check the validity of the condition (\ref{DKbound}) for our BH case
study to see if the minimal $\Delta _{k}$\ gives the most accurate WKB
order, and results are given in Tables \ref{tab9}-\ref{tab12}. The minimal $%
\Delta _{k}$\ and $\delta _{k}$\ are denoted in bold style. By considering these tables, we find that the condition (\ref%
{DKbound}) is valid for LQBHs in all cases as for the Schwarzschild BHs and
conformal Weyl solutions. More interestingly, we see that to obtain the QN
frequencies by employing the higher-order WKB formula (\ref{wkb}), the
minimal $\Delta _{k}$ usually identifies the most accurate WKB order.

In addition, the QN modes are calculated through the various orders of Pad%
\'{e} approximation and results are presented in Tables \ref{tab13}-\ref%
{tab16}. The bold values denote the minimal SD\ and $\delta _{k}$. From
these tables, it is clear that the minimal SD coincides with the minimal $%
\delta _{k}$\ except for $n=0=l$. Thus, the minimum SD gives the most
accurate result that could be obtained through the Pad\'{e} approximants. On
the other hand, by comparing Tables \ref{tab9}-\ref{tab12} with Tables %
\ref{tab13}-\ref{tab16} in order, we see how Pad\'{e} approximants increase
the accuracy of the WKB formula. Therefore, even for the case $n=0=l$,
employing the Pad\'{e} approximation with minimal SD is more accurate than
the ordinary WKB approximation, as we expected.
\begin{center}
\begin{table*}[!htb]
\scalebox{1} {\begin{tabular}{|c|c|c|c|c|c|c|c|c|c|} \hline\hline
$k$ &  & $\omega _{k}\left( l=0\right) $ & $\Delta _{k}$ & $\delta
_{k}$ &
&  & $\omega _{k}\left( l=1\right) $ & $\Delta _{k}$ & $\delta _{k}$ \\
\hline\hline $1$ &  & $0.3813-0.2084i$ & $-$ & $0.1806$ &  &  &
$0.6398-0.1962i$ & $-$ & $0.0823$ \\ \hline $2$ &  &
$0.2474-0.3212i$ & $0.1219$ & $0.1245$ &  &  & $0.5574-0.2252i$ &
$0.0476$ & $0.0322$ \\ \hline $3$ &  & $0.1409-0.2487i$ & $0.0696$
& $0.0737$ &  &  & $0.5447-0.1917i$ & $0.0192$ & $0.0130$ \\
\hline $4$ &  & $0.1713-0.2046i$ & $\mathbf{0.0341}$ &
$\mathbf{0.0296}$ &  &  & $0.5590-0.1868i$ & $0.0087$ & $0.0064$
\\ \hline $5$ &  & $0.1416-0.1805i$ & $0.0530$ & $0.0643$ &  &  &
$0.5617-0.1948i$ & $\mathbf{0.0061}$ & $\mathbf{0.0045}$ \\ \hline
$6$ &  & $0.2254-0.1134i$ & $0.0712$ & $0.0955$ &  &  &
$0.5537-0.1977i$ & $0.0085$ & $0.0061$ \\ \hline $7$ &  &
$0.2821-0.2041i$ & $0.0810$ & $0.0813$ &  &  & $0.5488-0.1837i$ &
$0.0168$ & $0.0128$ \\ \hline $8$ &  & $0.2097-0.2746i$ & $0.0994$
& $0.0695$ &  &  & $0.5780-0.1744i$ & $0.0349$ & $0.0276$ \\
\hline $9$ &  & $0.3488-0.3913i$ & $0.6479$ & $0.2374$ &  &  &
$0.5983-0.2330i$ & $0.0751$ & $0.0571$ \\ \hline $10$ &  &
$0.0872-1.5646i$ & $>1$ & $>1$ &  &  & $0.4815-0.2896i$ & $0.1527$
& $0.1230$ \\ \hline $11$ &  & $2.0551+2.5815i$ & $>1$ & $>1$ &  &
& $0.3850-0.0146i$ & $0.1504$ & $0.2483$ \\ \hline $12$ &  &
$2.4757+2.1429i$ & $>1$ & $>1$ &  &  & $0.5914-0.0095i$ & $0.5055$
& $0.1866$ \\ \hline $13$ &  & $11.41+11.35i$ & $-$ & $>1$ &  &  &
$0.9907+0.7949i$ & $-$ & $>1$
\\ \hline\hline
\end{tabular}}
\caption{The QN modes calculated by the WKB formula of different orders for $%
P=0$, $A_{0}=0.5$, $n=0$, $l=0$ (left), and $l=1$ (right). The
minimum value of $\Delta _{k}$ and $\protect\delta _{k}$\ are
given in bold. The
accurate modes $\protect\omega _{00}=0.2008-0.2057i$\ and $\protect\omega %
_{01}=0.5576-0.1930i $\ are taken from Table \protect\ref{tab3}.}
\label{tab9}
\bigskip
\scalebox{1} {\begin{tabular}{|c|c|c|c|c|c|c|c|c|c|} \hline\hline
$k$ &  & $\omega _{k}\left( l=1\right) $ & $\Delta _{k}$ & $\delta
_{k}$ &
&  & $\omega _{k}\left( l=2\right) $ & $\Delta _{k}$ & $\delta _{k}$ \\
\hline\hline $1$ &  & $0.7780-0.4841i$ & $-$ & $0.3292$ &  &  &
$1.0872-0.5181i$ & $-$ & $0.2269$ \\ \hline $2$ &  &
$0.5385-0.6994i$ & $0.1858$ & $0.1195$ &  &  & $0.8906-0.6324i$ &
$0.1194$ & $0.0541$ \\ \hline $3$ &  & $0.4328-0.6216i$ & $0.0701$
& $0.0426$ &  &  & $0.8581-0.5858i$ & $0.0290$ & $0.0120$ \\
\hline $4$ &  & $0.4628-0.5814i$ & $0.0292$ & $0.0213$ &  &  &
$0.8686-0.5786i$ & $0.0067$ & $0.0040$ \\ \hline $5$ &  &
$0.4876-0.6014i$ & $\mathbf{0.0230}$ & $\mathbf{0.0173}$ &  &  &
$0.8712-0.5824i$ & $0.0025$ & $0.0016$ \\ \hline $6$ &  &
$0.4676-0.6271i$ & $0.0319$ & $0.0259$ &  &  & $0.8694-0.5836i$ &
$\mathbf{0.0014}$ & $\mathbf{0.0011}$ \\ \hline $7$ &  &
$0.4241-0.5953i$ & $0.0533$ & $0.0465$ &  &  & $0.8684-0.5820i$ &
$0.0017$ & $0.0013$ \\ \hline $8$ &  & $0.4840-0.5217i$ & $0.0742$
& $0.0807$ &  &  & $0.8708-0.5804i$ & $0.0032$ & $0.0025$ \\
\hline $9$ &  & $0.5722-0.6045i$ & $0.0605$ & $0.1020$ &  &  &
$0.8739-0.5850i$ & $0.0064$ & $0.0050$ \\ \hline $10$ &  &
$0.5735-0.6031i$ & $0.3579$ & $0.1032$ &  &  & $0.8645-0.5914i$ &
$0.0133$ & $0.0102$ \\ \hline $11$ &  & $1.0865-1.1024i$ & $>1$ &
$0.7943$ &  &  & $0.8510-0.5715i$ & $0.0286$ & $0.0216$ \\ \hline
$12$ &  & $0.3583-3.3435i$ & $>1$ & $>1$ &  &  & $0.8953-0.5432i$
& $0.0637$ & $0.0469$ \\ \hline $13$ &  & $4.6774+5.7383i$ & $-$ &
$>1$ &  &  & $0.9577-0.6409i$ & $-$ & $0.1057$ \\ \hline\hline
\end{tabular}}
\caption{The QN modes calculated by the WKB formula of different orders for $%
P=0$, $A_{0}=0.5$, $n=1$, $l=1$ (left), and $l=2$ (right). The
minimum value of $\Delta _{k}$ and $\protect\delta _{k}$\ are
given in bold. The
accurate modes $\protect\omega _{11}=0.4703-0.6013i$\ and $\protect\omega %
_{12}=0.8696-0.5825i $\ are taken from Table \protect\ref{tab4}.}
\label{tab10}
\bigskip
\begin{tabular}{|c|c|c|c|c|c|c|c|c|c|}
\hline\hline $k$ &  & $\omega _{k}\left( l=0\right) $ & $\Delta
_{k}$ & $\delta _{k}$ &
&  & $\omega _{k}\left( l=1\right) $ & $\Delta _{k}$ & $\delta _{k}$ \\
\hline\hline
$1$ &  & $0.3948-0.2043i$ & $-$ & $0.1656$ &  &  & $0.6920-0.2009i$ & $-$ & $%
>10^{-3}$ \\ \hline
$2$ &  & $0.2754-0.2930i$ & $0.0903$ & $0.0876$ &  &  & $0.6197-0.2244i$ & $%
>10^{-3}$ & $>10^{-3}$ \\ \hline
$3$ &  & $0.2177-0.2396i$ & $0.0399$ & $0.0246$ &  &  & $0.6128-0.2045i$ & $%
>10^{-3}$ & $>10^{-3}$ \\ \hline
$4$ &  & $0.2280-0.2288i$ & $0.0099$ & $0.0107$ &  &  & $0.6165-0.2033i$ & $%
>10^{-3}$ & $0.00051$ \\ \hline
$5$ &  & $0.2191-0.2199i$ & $0.0096$ & $0.0108$ &  &  & $0.6168-0.2039i$ & $%
0.0003713$ & $0.00029$ \\ \hline
$6$ &  & $0.2298-0.2097i$ & $0.0082$ & $0.0085$ &  &  & $0.6165-0.2040i$ & $%
0.0001834$ & $0.00024$ \\ \hline
$7$ &  & $0.2349-0.2153i$ & $\mathbf{0.0061}$ & $\mathbf{0.0059}$ &  &  & $%
0.6164-0.2038i$ & $0.0001122$ & $0.00011$ \\ \hline
$8$ &  & $0.2420-0.2090i$ & $0.0194$ & $0.0153$ &  &  & $0.6165-0.2038i$ & $%
\mathbf{0.0000462}$ & $\mathbf{<5\times 10}^{-5}$ \\ \hline
$9$ &  & $0.2669-0.2373i$ & $0.0298$ & $0.0417$ &  &  & $0.6165-0.2038i$ & $%
0.0000610$ & $\mathbf{<5\times 10}^{-5}$ \\ \hline
$10$ &  & $0.2361-0.2682i$ & $0.0436$ & $0.0504$ &  &  & $0.6166-0.2038i$ & $%
0.0001349$ & $0.00011$ \\ \hline
$11$ &  & $0.2938-0.3202i$ & $0.1872$ & $0.1204$ &  &  & $0.6167-0.2040i$ & $%
0.0002370$ & $0.00025$ \\ \hline
$12$ &  & $0.1488-0.6323i$ & $0.4874$ & $0.4220$ &  &  & $0.6163-0.2041i$ & $%
0.0003873$ & $0.00038$ \\ \hline $13$ &  & $0.2165+0.6516i$ & $-$
& $0.8699$ &  &  & $0.6161-0.2035i$ & $-$ & $0.00052$ \\
\hline\hline
\end{tabular}%
\caption{The QN modes calculated by the WKB formula of different orders for $%
A_{0}=0$, $P=0.02$, $n=0$, $l=0$ (left), and $l=1$ (right). The
minimum value of $\Delta _{k}$ and $\protect\delta _{k}$\ are
given in bold.
The accurate modes $\protect\omega _{00}=0.2298-0.2182i$\ and $\protect%
\omega _{01}=0.6165-0.2038i $\ are taken from Table
\protect\ref{tab6}.} \label{tab11}
\end{table*}
\end{center}
\begin{center}
\begin{table*}[!htb]
\begin{tabular}{|c|c|c|c|c|c|c|c|c|c|}
\hline\hline $k$ &  & $\omega _{k}\left( l=1\right) $ & $\Delta
_{k}$ & $\delta _{k}$ &
&  & $\omega _{k}\left( l=2\right) $ & $\Delta _{k}$ & $\delta _{k}$ \\
\hline\hline $1$ &  & $0.8310-0.5019i$ & $-$ & $>10^{-2}$ &  &  &
$1.1798-0.5442i$ & $-$ & $>10^{-2}$ \\ \hline $2$ &  &
$0.6067-0.6875i$ & $>10^{-2}$ & $>10^{-2}$ &  &  &
$0.9949-0.6453i$ & $>10^{-2}$ & $>10^{-2}$ \\ \hline $3$ &  &
$0.5536-0.6410i$ & $>10^{-2}$ & $0.00501$ &  &  & $0.9771-0.6176i$
& $>10^{-2}$ & $0.00142$ \\ \hline $4$ &  & $0.5564-0.6378i$ &
$0.002423$ & $0.00184$ &  &  & $0.9783-0.6169i$ & $0.0007025$ &
$0.00041$ \\ \hline $5$ &  & $0.5581-0.6392i$ & $0.001103$ &
$0.00054$ &  &  & $0.9785-0.6172i$ & $0.0002033$ &
$\mathbf{<5\times 10}^{-5}$ \\ \hline
$6$ &  & $0.5581-0.6392i$ & $\mathbf{0.000168}$ & $0.00050$ &  &  & $%
0.9785-0.6172i$ & $0.0000281$ & $\mathbf{<5\times 10}^{-5}$ \\
\hline $7$ &  & $0.5579-0.6390i$ & $0.000189$ & $0.00031$ &  &  &
$0.9784-0.6172i$ & $0.0000297$ & $0.00007$ \\ \hline
$8$ &  & $0.5580-0.6388i$ & $0.000205$ & $\mathbf{0.00014}$ &  &  & $%
0.9785-0.6172i$ & $0.0000196$ & $\mathbf{<5\times 10}^{-5}$ \\
\hline $9$ &  & $0.5577-0.6386i$ & $0.000399$ & $0.00032$ &  &  &
$0.9785-0.6172i$ & $0.0000082$ & $\mathbf{<5\times 10}^{-5}$ \\
\hline $10$ &  & $0.5582-0.6381i$ & $0.000728$ & $0.00066$ &  &  &
$0.9785-0.6172i$ & $0.0000048$ & $\mathbf{<5\times 10}^{-5}$ \\
\hline $11$ &  & $0.5591-0.6389i$ & $0.001194$ & $0.00114$ &  &  &
$0.9785-0.6172i$ & $0.0000029$ & $\mathbf{<5\times 10}^{-5}$ \\
\hline $12$ &  & $0.5578-0.6404i$ & $0.001858$ & $0.00173$ &  &  &
$0.9785-0.6172i$ & $\mathbf{0.0000021}$ & $\mathbf{<5\times
10}^{-5}$ \\ \hline $13$ &  & $0.5554-0.6384i$ & $-$ & $0.00258$ &
&  & $0.9785-0.6172i$ & $-$ & $\mathbf{<5\times 10}^{-5}$ \\
\hline\hline
\end{tabular}%
\caption{The QN modes calculated by the WKB formula of different orders for $%
A_{0}=0$, $P=0.02$, $n=1$, $l=1$ (left), and $l=2$ (right). The
minimum value of $\Delta _{k}$ and $\protect\delta _{k}$\ are
given in bold.
The accurate modes $\protect\omega _{11}=0.5580-0.6387i$\ and $\protect%
\omega _{12}=0.9785-0.6172i $\ are taken from Table
\protect\ref{tab7}.} \label{tab12}
\bigskip
\begin{tabular}{|c|c|c|c|c|c|c|c|c|c|}
\hline\hline
$k$ &  & $\omega _{k}\left( l=0\right) $ & SD & $\delta _{k}$ &  &  & $%
\omega _{k}\left( l=1\right) $ & SD & $\delta _{k}$ \\
\hline\hline $1$ &  & $0.2937-0.1605i$ & $>10^{-2}$ & $>10^{-2}$ &
&  & $0.5848-0.1794i$ & $>10^{-2}$ & $>10^{-2}$ \\ \hline $2$ &  &
$0.2089-0.1827i$ & $>10^{-2}$ & $>10^{-2}$ &  &  &
$0.5563-0.1913i$ & $0.005019$ & $0.00214$ \\ \hline $3$ &  &
$0.1982-0.1962i$ & $0.003302$ & $0.00985$ &  &  & $0.5571-0.1918i$
& $0.000329$ & $0.00126$ \\ \hline
$4$ &  & $0.1989-0.1965i$ & $\mathbf{0.000304}$ & $0.00937$ &  &  & $%
0.5582-0.1928i$ & $0.000582$ & $0.00067$ \\ \hline $5$ &  &
$0.2012-0.1965i$ & $0.003240$ & $0.00917$ &  &  & $0.5571-0.1932i$
& $0.000337$ & $0.00056$ \\ \hline $6$ &  & $0.2040-0.2001i$ &
$0.000858$ & $0.00647$ &  &  & $0.5576-0.1928i$ & $0.000188$ &
$0.00022$ \\ \hline $7$ &  & $0.2037-0.1996i$ & $0.000387$ &
$0.00677$ &  &  & $0.5576-0.1928i$ & $0.000142$ & $0.00015$ \\
\hline $8$ &  & $0.2046-0.2002i$ & $0.001148$ & $0.00669$ &  &  &
$0.5577-0.1929i$ & $0.000068$ & $0.00014$ \\ \hline $9$ &  &
$0.2036-0.1995i$ & $0.000410$ & $0.00682$ &  &  & $0.5577-0.1930i$
& $0.000178$ & $0.00014$ \\ \hline $10$ &  & $0.2046-0.2006i$ &
$0.001567$ & $0.00630$ &  &  & $0.5577-0.1930i$ & $0.000023$ &
$0.00008$ \\ \hline $11$ &  & $0.2076-0.1967i$ & $0.008442$ &
$>10^{-2}$ &  &  & $0.5577-0.1930i$ & $0.000048$ & $0.00007$ \\
\hline
$12$ &  & $0.1996-0.2054i$ & $0.004183$ & $\mathbf{0.00125}$ &  &  & $%
0.5577-0.1930i$ & $0.000074$ & $0.00007$ \\ \hline $13$ &  &
$0.2019-0.2083i$ & $0.009697$ & $0.00281$ &  &  & $0.5576-0.1930i$
& $\mathbf{0.000007}$ & $\mathbf{<5\times 10}^{-5}$ \\
\hline\hline
\end{tabular}%
\caption{The QN modes calculated by averaging of Pad\'{e}
approximations of different orders for $P=0$, $A_{0}=0.5$, $n=0$,
$l=0$ (left), and $l=1$ (right). The minimal SD and
$\protect\delta _{k}$\ are given in bold.
The accurate modes $\protect\omega _{00}=0.2008-0.2057i$\ and $\protect%
\omega _{01}=0.5576-0.1930i$\ are taken from Table
\protect\ref{tab3}.} \label{tab13}
\bigskip
\begin{tabular}{|c|c|c|c|c|c|c|c|c|c|}
\hline\hline
$k$ &  & $\omega _{k}\left( l=1\right) $ & SD & $\delta _{k}$ &  &  & $%
\omega _{k}\left( l=2\right) $ & SD & $\delta _{k}$ \\
\hline\hline $1$ &  & $0.5608-0.3490i$ & $>10^{-2}$ & $>10^{-2}$ &
&  & $0.8860-0.4222i$ & $>10^{-2}$ & $>10^{-2}$ \\ \hline $2$ &  &
$0.5161-0.5923i$ & $>10^{-2}$ & $>10^{-2}$ &  &  &
$0.8829-0.5814i$ & $>10^{-2}$ & $>10^{-2}$ \\ \hline $3$ &  &
$0.4539-0.6082i$ & $>10^{-2}$ & $>10^{-2}$ &  &  &
$0.8642-0.5834i$ & $0.006589$ & $0.00543$ \\ \hline $4$ &  &
$0.4694-0.5918i$ & $0.007131$ & $0.00952$ &  &  & $0.8694-0.5805i$
& $0.001170$ & $0.00198$ \\ \hline $5$ &  & $0.4731-0.6019i$ &
$0.006718$ & $0.00285$ &  &  & $0.8702-0.5823i$ & $0.000844$ &
$0.00062$ \\ \hline $6$ &  & $0.4661-0.6026i$ & $0.001203$ &
$0.00443$ &  &  & $0.8693-0.5828i$ & $0.000094$ & $0.00042$ \\
\hline $7$ &  & $0.4715-0.5943i$ & $>10^{-2}$ & $0.00708$ &  &  &
$0.8696-0.5822i$ & $0.000386$ & $0.00031$ \\ \hline $8$ &  &
$0.4702-0.6005i$ & $0.000487$ & $0.00083$ &  &  & $0.8697-0.5826i$
& $0.000032$ & $0.00013$ \\ \hline $9$ &  & $0.4726-0.6115i$ &
$>10^{-2}$ & $>10^{-2}$ &  &  & $0.8696-0.5826i$ & $0.000031$ &
$0.00007$ \\ \hline $10$ &  & $0.4965-0.6019i$ & $>10^{-2}$ &
$>10^{-2}$ &  &  & $0.8696-0.5825i$ & $0.000011$ &
$\mathbf{<5\times 10}^{-5}$ \\ \hline $11$ &  & $0.4703-0.6017i$ &
$0.000198$ & $0.00037$ &  &  & $0.8696-0.5826i$ & $0.000027$ &
$0.00007$ \\ \hline $12$ &  & $0.4703-0.6014i$ &
$\mathbf{0.000174}$ & $\mathbf{0.00011}$ &  & & $0.8696-0.5825i$ &
$0.000010$ & $\mathbf{<5\times 10}^{-5}$ \\ \hline $13$ &  &
$0.4698-0.6015i$ & $0.000338$ & $0.00051$ &  &  & $0.8696-0.5825i$
& $\mathbf{0.000008}$ & $\mathbf{<5\times 10}^{-5}$ \\
\hline\hline
\end{tabular}%
\caption{The QN modes calculated by averaging of Pad\'{e}
approximations of different orders for $P=0$, $A_{0}=0.5$, $n=1$,
$l=1$ (left), and $l=2$ (right). The minimal SD and
$\protect\delta _{k}$\ are given in bold.
The accurate modes $\protect\omega _{11}=0.4703-0.6013i$\ and $\protect%
\omega _{12}=0.8696-0.5825i $\ are taken from Table
\protect\ref{tab4}.} \label{tab14}
\end{table*}
\end{center}
\begin{center}
\begin{table*}[!htb]
\begin{tabular}{|c|c|c|c|c|c|c|c|c|c|}
\hline\hline
$k$ &  & $\omega _{k}\left( l=0\right) $ & SD & $\delta _{k}$ &  &  & $%
\omega _{k}\left( l=1\right) $ & SD & $\delta _{k}$ \\
\hline\hline $1$ &  & $0.3114-0.1612i$ & $>10^{-2}$ & $>10^{-2}$ &
&  & $0.6382-0.1853i$ & $>10^{-4}$ & $>10^{-3}$ \\ \hline $2$ &  &
$0.2375-0.1917i$ & $>10^{-2}$ & $>10^{-2}$ &  &  &
$0.6166-0.2010i$ & $>10^{-4}$ & $>10^{-3}$ \\ \hline $3$ &  &
$0.2309-0.2085i$ & $0.004396$ & $0.00974$ &  &  & $0.6161-0.2032i$
& $>10^{-4}$ & $0.00075$ \\ \hline $4$ &  & $0.2284-0.2220i$ &
$0.007874$ & $0.00403$ &  &  & $0.6166-0.2038i$ & $>10^{-4}$ &
$0.00007$ \\ \hline $5$ &  & $0.2314-0.2156i$ & $0.001599$ &
$0.00303$ &  &  & $0.6166-0.2039i$ & $0.00008569$ & $0.00013$ \\
\hline $6$ &  & $0.2316-0.2175i$ & $0.001262$ & $0.00192$ &  &  &
$0.6165-0.2038i$ & $0.00000082$ & $\mathbf{<5\times 10}^{-5}$ \\
\hline $7$ &  & $0.2328-0.2171i$ & $0.000709$ & $0.00324$ &  &  &
$0.6165-0.2038i$ & $\mathbf{0.00000001}$ & $\mathbf{<5\times
10}^{-5}$ \\ \hline $8$ &  & $0.2305-0.2180i$ & $0.000175$ &
$0.00071$ &  &  & $0.6165-0.2038i$ & $0.00000094$ &
$\mathbf{<5\times 10}^{-5}$ \\ \hline
$9$ &  & $0.2295-0.2183i$ & $0.001963$ & $\mathbf{0.00032}$ &  &  & $%
0.6165-0.2038i$ & $0.00002807$ & $\mathbf{<5\times 10}^{-5}$ \\
\hline $10$ &  & $0.2300-0.2178i$ & $0.000837$ & $0.00047$ &  &  &
$0.6165-0.2038i$ & $0.00000233$ & $\mathbf{<5\times 10}^{-5}$ \\
\hline
$11$ &  & $0.2303-0.2181i$ & $\mathbf{0.000675}$ & $0.00053$ &  &  & $%
0.6165-0.2038i$ & $0.00000018$ & $\mathbf{<5\times 10}^{-5}$ \\
\hline $12$ &  & $0.2310-0.2181i$ & $0.001046$ & $0.00121$ &  &  &
$0.6165-0.2038i$ & $0.00000004$ & $\mathbf{<5\times 10}^{-5}$ \\
\hline $13$ &  & $0.2302-0.2183i$ & $0.000868$ & $0.00043$ &  &  &
$0.6165-0.2038i$ & $0.00000008$ & $\mathbf{<5\times 10}^{-5}$ \\
\hline\hline
\end{tabular}%
\caption{The QN modes calculated by averaging of Pad\'{e}
approximations of different orders for $A_{0}=0$, $P=0.02$, $n=0$,
$l=0$ (left), and $l=1$ (right). The minimal SD and
$\protect\delta _{k}$\ are given in bold.
The accurate modes $\protect\omega _{00}=0.2298-0.2182i$\ and $\protect%
\omega _{01}=0.6165-0.2038i $\ are taken from Table
\protect\ref{tab6}.} \label{tab15}
\bigskip
\begin{tabular}{|c|c|c|c|c|c|c|c|c|c|}
\hline\hline
$k$ &  & $\omega _{k}\left( l=1\right) $ & SD & $\delta _{k}$ &  &  & $%
\omega _{k}\left( l=2\right) $ & SD & $\delta _{k}$ \\
\hline\hline $1$ &  & $0.6088-0.3678i$ & $>10^{-2}$ & $>10^{-2}$ &
&  & $0.9729-0.4487i$ & $>10^{-4}$ & $>10^{-3}$ \\ \hline $2$ &  &
$0.5875-0.6015i$ & $>10^{-2}$ & $>10^{-2}$ &  &  &
$0.9881-0.6033i$ & $>10^{-4}$ & $>10^{-3}$ \\ \hline $3$ &  &
$0.5602-0.6363i$ & $0.008200$ & $0.00330$ &  &  & $0.9794-0.6166i$
& $>10^{-4}$ & $>10^{-3}$ \\ \hline $4$ &  & $0.5562-0.6372i$ &
$0.001406$ & $0.00238$ &  &  & $0.9782-0.6167i$ & $>10^{-4}$ &
$0.00052$ \\ \hline $5$ &  & $0.5576-0.6394i$ & $0.000505$ &
$0.00081$ &  &  & $0.9784-0.6172i$ & $0.00005898$ & $0.00008$ \\
\hline $6$ &  & $0.5581-0.6392i$ & $0.000040$ & $0.00047$ &  &  &
$0.9785-0.6172i$ & $0.00000197$ & $\mathbf{<5\times 10}^{-5}$ \\
\hline $7$ &  & $0.5579-0.6391i$ & $0.000116$ & $0.00042$ &  &  &
$0.9784-0.6172i$ & $0.00001819$ & $0.00006$ \\ \hline $8$ &  &
$0.5580-0.6390i$ & $0.000030$ & $0.00027$ &  &  & $0.9785-0.6172i$
& $0.00000141$ & $\mathbf{<5\times 10}^{-5}$ \\ \hline $9$ &  &
$0.5577-0.6387i$ & $0.000187$ & $0.00026$ &  &  & $0.9785-0.6172i$
& $0.00000410$ & $\mathbf{<5\times 10}^{-5}$ \\ \hline $10$ &  &
$0.5582-0.6389i$ & $0.000331$ & $0.00029$ &  &  & $0.9785-0.6172i$
& $0.00000112$ & $\mathbf{<5\times 10}^{-5}$ \\ \hline $11$ &  &
$0.5581-0.6388i$ & $0.000188$ & $0.00010$ &  &  & $0.9785-0.6172i$
& $0.00000010$ & $\mathbf{<5\times 10}^{-5}$ \\ \hline $12$ &  &
$0.5580-0.6387i$ & $0.000019$ & $\mathbf{<5\times 10}^{-5}$ &  & &
$0.9785-0.6172i$ & $0.00000043$ & $\mathbf{<5\times 10}^{-5}$ \\
\hline
$13$ &  & $0.5580-0.6387i$ & $\mathbf{0.000008}$ & $\mathbf{<5\times 10}%
^{-5} $ &  &  & $0.9785-0.6172i$ & $\mathbf{0.00000007}$ &
$\mathbf{<5\times 10}^{-5}$ \\ \hline\hline
\end{tabular}%
\caption{The QN modes calculated by averaging of Pad\'{e}
approximations of different orders for $A_{0}=0$, $P=0.02$, $n=1$,
$l=1$ (left), and $l=2$ (right). The minimal SD and
$\protect\delta _{k}$\ are given in bold.
The accurate modes $\protect\omega _{11}=0.5580-0.6387i$\ and $\protect\omega %
_{12}=0.9785-0.6172i $\ are taken from Table \protect\ref{tab7}.}
\label{tab16}
\end{table*}
\end{center}

\clearpage

\section{Conclusions \label{Conclusions}}

We have considered a minimally coupled scalar perturbation in the background
spacetime of the LQG-corrected BHs characterized by two LQG correction
parameters, namely, the polymeric function $P$ and the minimum area gap $%
A_{0}$. We have calculated the corresponding QN modes with the help of three
independent methods of calculations; the higher-order WKB formula and
related Pad\'{e} approximants, the improved AIM, and time-domain
integration. The effects of LQG correction parameters on the QNMs spectrum
have been studied and deviations from those of the Schwarzschild BHs have
been investigated.

We have found that the QNMs were more sensitive to changes in the polymeric
function $P$ compared with the minimum area gap $A_{0}$. Thus, $P$ plays a
more important role in the evolution of fields on the background geometry of
LQBHs compared with $A_{0}$. In addition, we have shown that the LQG
correction parameters had opposite effects on the QN frequencies. Increasing
in $P$\ ($A_{0}$) led to increasing (decreasing) in the real part of
frequencies and damping rate. While one of the free parameters increases the
lifetime of perturbations, the other one attempts to dissipate perturbations
faster. These cases have been also confirmed through the time-domain profile
of perturbations by considering the contribution of all modes. We have also
calculated the dominant QN frequencies by employing the Prony method which
was in good agreement with the results of AIM.

In addition, we have seen that the effective potential of perturbations was
positive definite and the modes decayed in time that guaranteed the
dynamical stability of the LQBHs undergoing scalar perturbations. Although a
negative gap appeared in the effective potential for the lowest multipole
number and higher values of the LQG correction parameters, the perturbations
decayed in time which indicated dynamical stability of the BHs.

We have used the higher-order WKB formula and related Pad\'{e}
approximants as a semi-analytic method to obtain the QNMs and find
the most accurate order of the WKB and Pad\'{e} approximations for
calculating the QN frequencies. It was shown that the minimum
value of error estimation quantity, denoted by $\Delta _{k}$
throughout the text, provides a good estimation for the error and usually gives the most accurate WKB order.
Besides, we have seen that by employing the averaging of Pad\'{e}
approximations, one can increase the accuracy of modes
considerably compared to the ordinary WKB formula and obtain
accurate modes for $n<l$.


\section*{Acknowledgements}

The author is grateful to FORDECYT-PRONACES-CONACYT for support under
Grant No. CF-MG-2558591. He also acknowledges financial assistance from CONACYT through the postdoctoral Grant No. 31155.


\end{document}